\documentclass[journal]{IEEEtran}

\usepackage{xcolor,soul,framed} 

\colorlet{shadecolor}{yellow}
\usepackage[pdftex]{graphicx}
\graphicspath{{../pdf/}{../jpeg/}}
\DeclareGraphicsExtensions{.pdf,.jpeg,.png}

\usepackage[cmex10]{amsmath}
\usepackage{mathrsfs}
\usepackage{bbm}
\usepackage{array}
\usepackage{mdwmath}
\usepackage{mdwtab}
\usepackage{eqparbox}
\usepackage{url}
\usepackage{algorithm}
\usepackage{algorithmic}
\usepackage{multirow}
\usepackage{amsmath}
\usepackage{hhline}
\usepackage{makecell}
\usepackage{color}
\usepackage[hidelinks]{hyperref}
\usepackage{amsfonts,amssymb}
\usepackage{subcaption}
\usepackage[labelsep=period, font=small]{caption}

\hypersetup{
    colorlinks=true,
    citecolor=blue,
    linkcolor=black
}
\usepackage{cite}

\begin{document}
\bstctlcite{IEEEexample:BSTcontrol}
    \title{Privacy-Aware Spectrum Pricing and Power Control Optimization for LEO Satellite Internet-of-Things}

\author{Bowen Shen,
        Kwok-Yan Lam,~\IEEEmembership{Senior Member,~IEEE}, 
        Feng Li,~\IEEEmembership{Member,~IEEE}
        and Li Wang


\thanks{B. Shen and K. Lam are with the School of Computer Science and Engineering, Nanyang Technological University, 639798, Singapore. (bowen010@e.ntu.edu.sg, kwokyan.lam@ntu.edu.sg)}

\thanks{F. Li is with the School of Information and Electronic Engineering, Zhejiang Gongshang University, Hangzhou, 310018, China, and also with the Strategic Centre for Research in Privacy-Preserving Technologies and Systems, Nanyang Technological University, 639798, Singapore. (li\_feng@ntu.edu.sg)}

\thanks{L. Wang is with the College of Marine Electrical Engineering, Dalian Maritime University, Dalian, 116026, China. (liwang2002@dlmu.edu.cn)}

}

\markboth{}{Roberg \MakeLowercase{\textit{}}: High-Efficiency Diode and Transistor Rectifiers}

\maketitle

\begin{abstract}

Low Earth orbit (LEO) satellite systems play an important role in next generation communication networks due to their ability to provide extensive global coverage with guaranteed communications in remote areas and isolated areas where base stations cannot be cost-efficiently deployed. With the pervasive adoption of LEO satellite systems, especially in the LEO Internet-of-Things (IoT) scenarios, their spectrum resource management requirements have become more complex as a result of massive service requests and high bandwidth demand from terrestrial terminals. For instance, when leasing the spectrum to terrestrial users and controlling the uplink transmit power, satellites collect user data for machine learning purposes, which usually are sensitive information such as location, budget and quality of service (QoS) requirement. To facilitate model training in LEO IoT while preserving the privacy of data, blockchain-driven federated learning (FL) is widely used by leveraging on a fully decentralized architecture. In this paper, we propose a hybrid spectrum pricing and power control framework for LEO IoT by combining blockchain technology and FL. We first design a local deep reinforcement learning algorithm for LEO satellite systems to learn a revenue-maximizing pricing scheme. Then the agents collaborate to form an FL system. We also propose a reputation-based blockchain which is used in the global model aggregation phase of FL to optimize the power control. Based on the reputation mechanism, a node is selected for each global training round to perform model aggregation and block generation, which can further enhance the decentralization of the network and guarantee the trust. Simulation tests are conducted to evaluate the performances of the proposed scheme. Our results show the efficiency of finding the maximum revenue scheme for LEO satellite systems while preserving the privacy of each agent.


\end{abstract}


%
\IEEEpeerreviewmaketitle


\section{Introduction}

\IEEEPARstart{F}{or} many regions such as oceans and deserts, which account for most of the Earth's surface, it is not easy to deploy massive base stations (BSs) to support continuously upgrading wireless demands and massive Internet-of-Things (IoT) terminals \cite{Ding2023, Yao2018}. As the extension of terrestrial BSs, satellite communications especially low Earth orbit (LEO) satellite communications have attracted many researchers' and practitioners' interest due to the advantage of highly global coverage and guaranteed communications \cite{Lakew2023}. Many satellite operators such as Starlink, OneWeb, Amazon and Boeing have launched or are planning to launch LEO satellite networks to cover millions of potential terrestrial terminals \cite{Starlink, OneWeb, Boeing}. The low orbital altitude of the satellite makes the transmission delay shorter and the path loss smaller compared to geostationary Earth orbit (GEO) satellites, and the constellation composed of multiple satellites can achieve global coverage. Besides, cellular communication, multiple access, point beam, frequency multiplexing and other technologies also provide the technical guarantee for LEO satellite communications. Sixth generation (6G) communications, which are built on the base of LEO satellite networks, will be driven by the surging artificial intelligence (AI), big data, and Internet of Everything (IoE) technologies.  In this context, the mobile networks of 6G and beyond are expected to not only enhance the key performance indicators and quality of service (QoS) of 5G continuously but also introduce numerous novel technologies and use cases \cite{Xuewen2024}.

With the blooming terrestrial IoT applications in recent years, existing wireless resources can not meet the requirements in many fields including vehicular communications, industrial automation, sensor networks, and public safety \cite{Gill2023, Chang2023, Safdar2023, Shafigh2023, Li5G, LiAgent}. As the demand for spectrum resources and the number of massive user access increases rapidly, how to manage spectrum resources efficiently has become a key challenge for LEO satellite communication networks \cite{Huang2022}. Many techniques including dynamic spectrum access (DSA), non-orthogonal multiple access (NOMA), cognitive radio (CR) and multiple spot beams have been proposed to alleviate the pressure on spectrum resource usage \cite{Li2020, Gu2022, Kaur2023, Qureshi2023}. In most scenarios when using DSA, deep learning is applied to train a model for resource allocation \cite{Yang2024}. Hence, satellites need to guarantee computing power for the model training. In \cite{Gao2020}, the authors combined the NOMA and orthogonal frequency division multiplexing to improve spectrum efficiency. In the utilization of cognitive radio (CR), wireless communication systems adaptively adjust their transmitting parameters by sensing the current communication environment. This adaptive approach enables efficient utilization of spectrum resources \cite{Qureshi2023}. The multiple spot beams technique can transfer a wide beam into multiple beams to increase the coverage gain of a satellite antenna, wherein interference between beams will affect the performance of the system \cite{Torkzaban2021}.

In recent years, self-learning-based methods, especially deep reinforcement learning (DRL) have become a focus in the field of DSA and spectrum sensing \cite{Cong2021, Yadav2022, yangh2024, LiDynamic}. Each user has a model that is regarded as an Agent that continually updates its parameters during training. The Agent interacts with the communication environment to find the optimal scheme. In order to enhance the cooperation among satellite nodes during training and to improve the training efficiency while still preserving the privacy of each node, federated learning (FL) and blockchain technology attract much attention \cite{Sun2023, Warnat2021, Passerat2020}. In the FL training process, each node's local model parameters instead of raw data are uploaded to the blockchain network. Then, a node is selected by the blockchain generation mechanism to conduct global model aggregation for the global training round. Such FL-based schemes improve the training efficiency of each node and also enhance the decentralization degree of the distributed system.

In this paper, based on blockchain-driven FL, we introduce a privacy-aware spectrum pricing and uplink transmit power control optimization scheme for LEO satellite IoT. Specifically, we first formulate the price bargaining between terrestrial users and LEO satellites as a Markov decision process. The service quality requirements of each terrestrial user and the condition of the satellites' spectrum change frequently. Deployment of reinforcement learning allows pricing and power control schemes to be adjusted in real time based on the changing environment. We use the Double Deep Q-learning to train a neural network model for each LEO satellite to find the optimal spectrum price. Besides, due to the limited battery capacity of LEO satellites, it is impractical to consume large amounts of power for model training on satellites. Thus, each LEO satellite has a terrestrial server for data computing. After receiving information from terrestrial users, the LEO satellite then sends it to the corresponding terrestrial server for model training. Considering different nodes have different computation power and the transaction information needs to be kept highly confidential, FL is applied in this paper for satellites' model training collaboration while privacy preservation is guaranteed to some degree. Traditional FL usually has a central server for global model aggregation and release. Thus, each node needs to give quite a lot of trust to the central server and the whole system will be paralyzed if the central server is malicious. To enhance the decentralization of the LEO satellite IoT networks, we introduce blockchain technology in the global model aggregation phase of FL. A reputation-based consensus mechanism is proposed based on the feature of transactions between terrestrial users and LEO satellites. Each LEO satellite that participates in the FL has a reputation record that determines the node to conduct the model aggregation and block generation in the global training round. And the behaviors of users who trade with the satellite will be used as the basis for increasing or decreasing the satellite's reputation.

The contributions of this paper can be highlighted as follows.
\begin{itemize}
    \item A reinforcement learning problem is formulated based on the Markov decision process to obtain an optimal policy for maximizing the revenues of LEO satellite systems by optimizing spectrum pricing.
    \item A DRL-based spectrum pricing scheme is proposed for LEO satellite IoT. We take into account the interference among terrestrial users in the same cell and try to find the optimal spectrum and power management scheme to balance the interference and maximize the benefits of LEO satellites.
    \item A blockchain-driven FL framework is designed. Based on the transaction characteristics between terrestrial users and LEO satellites, we introduce a reputation-based mechanism for the blockchain network to guarantee the suitability of global model aggregation and drive the LEO satellite to supervise and control the transmission power of terrestrial users.
    \item Simulations are conducted to evaluate the performance of the proposed framework. We present the performance of reward, price and revenue with different visibilities, and relative velocities of the satellites. We also considered the impact of the number of users and the performance comparisons of other methods.
\end{itemize}

The rest of this paper is organized as follows. Section II introduces the system model of the proposed scheme. In section III, we detail the scheme of the DRL-based spectrum pricing and power control. And the framework of blockchain-driven FL is also presented. Section IV shows the numerical results and section V concludes this paper finally.

\section{Related Work}
\begin{table}
    \caption{SUMMARY OF SYMBOLS AND NOTATIONS}
    \label{tab:Notations}
    \centering
    \begin{tabular}{|c|c|}
    \hline
         Symbols & Notations \\
         \hhline{|=|=|}
         $d_o^s$ & \makecell[c]{Distance between satellite and cell center} \\
         \hline
         $d_{Mn}^s$ & \makecell[c]{Distance between satellite and user $(M, n)$} \\
         \hline
         $R$ & \makecell[c]{Earth radius} \\
         \hline
         $d_{Mn}^o$ & \makecell[c]{Distance between cell center $o$ \\ and user $(M, n)$} \\
         \hline
         $\mathcal{P}_{n}$ & \makecell[c]{Transmit power of satellite terminal} \\
         \hline
         $\theta_n$ & \makecell[c]{Elevation angle from user $(M,n)$ to \\ the satellite system} \\
         \hline
         $g_{n}(\theta_n )$ & \makecell[c]{Antenna gain of user $(M,n)$ at the direction $\theta_n$} \\
         \hline
         $\alpha_{n}^M$ & \makecell[c]{Derivation angle form user $(M,n)$ to \\ the central line of cell $M$} \\
         \hline
         $G_{M}(\alpha_n^M)$ & \makecell[c]{Satellite antenna gain of cell $M$ at the direction $\alpha_n^M$} \\
         \hline
         $d_{n}$ & \makecell[c]{Straight-line distance between the user $(M,n)$  \\ and the satellite system} \\
         \hline
         $\lambda$ & \makecell[c]{wavelength} \\
         \hline
         $f_{n}(\theta_n)$ & \makecell[c]{Channel fading of user $(M, n)$ at the direction $\theta_n$} \\
         \hline
         $\mu_a$ & \makecell[c]{Active factor of user $a$ at cell $H$ which is related to \\ the user’s service type} \\
         \hline
         $\rho_H^M$ & \makecell[c]{Polarization isolation factor between cell $M$ and $H$} \\
         \hline
         $\upsilon^2$ & \makecell[c]{Power of noise} \\
         \hline
         $\mathrm{P}_s$ & \makecell[c]{Price of the LEO satellite's spectrum} \\
         \hline
         $C$ & \makecell[c]{Speed of light} \\
         \hline
         $\zeta$ & \makecell[c]{Revenue coefficient}\\
         \hline
         $F^d$ & \makecell[c]{Loss factor of Doppler shift} \\
         \hline
         $v_s$ & \makecell[c]{Relative velocity of satellite $s$} \\
         \hline
         $\gamma$ &  \makecell[c]{The angle between the direction of motion \\ and the direction of wave propagation}\\
         \hline
         $\varpi$ & \makecell[c]{Fading coefficient}\\
         \hline
         $\mathrm{B}_n$ & \makecell[c]{Budgets of terrestrial user $n$}\\
         \hline
         $u_{n}$ & \makecell[c]{Terrestrial users' utility} \\
         \hline
         $\mathcal{X}_n$ & \makecell[c]{Benefits obtained by contributing to the blockchain} \\
         \hline
         $\mathcal{S}$ & \makecell[c]{State} \\
         \hline
         $\mathcal{A}$ & \makecell[c]{Action} \\
         \hline
         $\mathcal{R}$ & \makecell[c]{Reward} \\
         \hline
         $Rep$ & \makecell[c]{Reputation token} \\
         \hline
         
    \end{tabular}
\end{table}

Many mathematical tools including Stackelberg game model have been widely explored to optimize spectrum resource utilization in satellite networks \cite{Li2023, Xie2023, POURK}. In \cite{Li2023}, the authors designed a multi-leader multi-follower Stackelberg game to achieve spectrum pricing. Seller operators, who are regarded as leaders, determine the pricing strategies based on the buying strategies of buyer operators who are regarded as followers. The authors defined the seller operations' revenue function as the income by providing bandwidth to buyer operators minus the service cost and charge for the primary node. And the buyer operations' revenue was expressed as an increasing function of the bought bandwidth. Then a Stackelberg game was formulated based on the two functions. In \cite{Xie2023}, the author formulated the problem of bandwidth pricing and allocation by employing a Stackelberg game-theoretic approach to model the interactions between spectrum providers and customers. Subsequently, the study analyzed the Stackelberg game equilibrium under two pricing strategies: uniform pricing and differential pricing. In the case of differential pricing, adjustments are made to individual customer prices based on various heterogeneous factors. In \cite{POURK}, game theory was used to model the wireless users' competition over shared spectrum. The author assumed that users who adjust a transmission power level to maximize their own utilities are players. And the utility of a player was evaluated based on the transmission rate. In \cite{bu}, the authors proposed a spectrum pricing method combined with blockchain technology. The spectrum pricing method takes advantage of the heterogeneity of LEO satellite spectrum by allowing a price differentiation between different spectrum ranges.

\begin{table*}[t]
\caption{Comparisons of Existing Spectrum Allocation and Power Control Optimization Schemes}
\centering
\begin{tabular}{@{\hspace{40pt}}c@{\hspace{40pt}}ccccc}
\hline
\rule{0pt}{10pt}Framework                                                                                                                 & \begin{tabular}[c]{@{}c@{}}Real-time Dynamic \\ Network Adaptation\end{tabular} & \begin{tabular}[c]{@{}c@{}}Interference Prevention\end{tabular} & \begin{tabular}[c]{@{}c@{}}Training Data \\ Privacy Preservation\end{tabular} & \begin{tabular}[c]{@{}c@{}}Resource Allocation \\ Auditability \end{tabular} & \begin{tabular}[c]{@{}c@{}}Satellite Mobility\end{tabular} \\ [3pt] \hline
\rule{0pt}{9pt}Ref. [29] & x & \checkmark & x & x & - \\ [1pt] \hline
\rule{0pt}{9pt}Ref. [30] & x & \checkmark & x & x & - \\ [1pt] \hline
\rule{0pt}{9pt}Ref. [32] & x & \checkmark & x & \checkmark & x \\ [1pt]\hline
\rule{0pt}{9pt}Ref. [33] & \checkmark & \checkmark & x & x & x \\ [1pt] \hline
\rule{0pt}{9pt}Ref. [37] & \checkmark & \checkmark & \checkmark & x & - \\ [1pt] \hline
\rule{0pt}{9pt}Ref. [38] & \checkmark & \checkmark & \checkmark & x & - \\ [1pt] \hline
\rule{0pt}{9pt}Ref. [40] & x & \checkmark & x & x & x \\ [1pt] \hline
\rule{0pt}{9pt}Ref. [42] & x & \checkmark & x & x & x \\ [1pt] \hline
\rule{0pt}{9pt}Ref. [44] & \checkmark & \checkmark & x & x & - \\ [1pt] \hline
\rule{0pt}{9pt}Our Solution & \checkmark & \checkmark & \checkmark & \checkmark & \checkmark \\ [1pt] \hline
\end{tabular}
\end{table*}

With the increasing computing power of mobile devices, the deployment of machine learning-based algorithms in satellite resource allocation attracts more attention \cite{Chan2022, Tang2022}. Satya Chan \emph{at al}. \cite{Chan2022} proposed a low complexity power and frequency resource allocation method to minimize inter-component interference while maximizing user throughput. This work first used a pre-trained perception to classify the condition of the traffic demand and then employed a projection tool to minimize the traffic demand reduction. Finally, a pre-trained linear regression model was introduced to allocate bandwidths. The scheme has excellent performance while keeping the low complexity of the algorithm. In \cite{Tang2022}, considering terrestrial users' limited battery capacity and each LEO satellite's computation capability, the authors trained a deep neural network model to minimize the total execution delay of terrestrial users.

In most satellite resource allocation scenarios, the complete information and environment conditions are generally difficult to get due to the dynamic environments. Hence, DRL has been adopted to address optimization problems in IoT networks. In \cite{Ahmed2024, Liao2020, XuewenDong2024}, the authors introduced the DRL methods in multibeam satellite systems for dynamic resource allocation. And multi-agent DRL scheme was proposed in \cite{Liao2020} to better address the cooperative game problems. Recently, there have been some studies about federated DRL (FDRL) for further collaborations between nodes in satellite IoT \cite{Liu2023, Yang2021, Yu2021}. In \cite{Liu2023}, the authors designed an adaptive FDRL scheme to find efficient task offloading and energy-saving policy considering the scenario of space-air-ground integrated edge computing. Considering the high communication costs and aggregation execution time, an asynchronous FL framework combined with a multi-agent asynchronous advantage actor-critic (A3C)-based joint device selection algorithm was proposed in \cite{Yang2021}. The scheme allows the users to update and aggregate the local model parameters asynchronously instead of waiting for devices with low computation powers. Besides, due to the A3C-based algorithm, the federated execution time and learning accuracy loss are effectively minimized. In \cite{Yu2021}, a two-timescale deep reinforcement learning (2Ts-DRL) approach, consisting of a fast-timescale and a slow-timescale learning process is proposed to achieve real-time and low overhead computation offloading decisions and resource allocation strategies in 5G ultra dense networks. TABLE II summarizes and compares the existing spectrum allocation and power control optimization schemes.

\begin{figure}[!t]
  \begin{center}
  \includegraphics[width=3.4in]{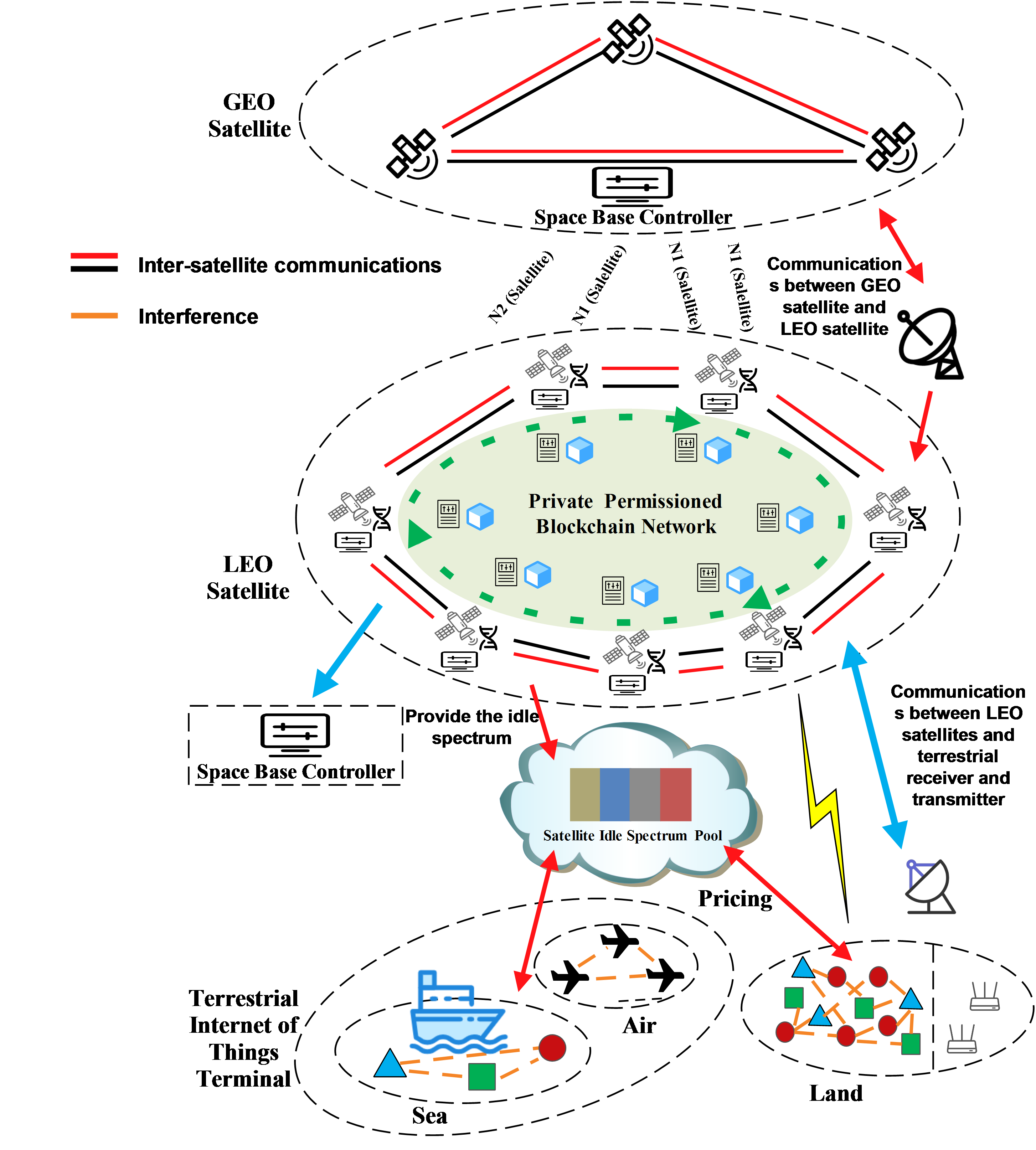}
  \caption{System model}\label{sim_opt_eff_classFinv}
  \end{center}
\end{figure}

\section{System Model}
\subsection {Satellite Network Architecture}

The spectrum pricing and sharing scheme in this paper is based on the LEO satellite IoT whose architecture is illustrated in Fig 1. It is assumed that the satellite payload is equipped with necessary modules such as multi-port amplifiers, flexible traveling wave tube amplifiers, etc. The scenario considered is that IoT nodes of terrestrial users are connected to LEO satellites through terrestrial cluster heads or BS. The terrestrial users carry transceivers compatible with both cellular and satellite data transmissions so that the cluster heads can communicate with the cluster members. LEO satellites share or lease their idle spectrum to terrestrial users directly or with the assistance of the GEO satellites to improve the utilization of the spectrum and increase their revenues. Each satellite has a terrestrial server for data computing and machine learning. Besides, these terrestrial servers are responsible for participating in FL for model training collaboration and a reputation-based blockchain due to the privacy preservation concern. During the process, based on the needs of the cluster members, the cluster head responds as a transaction agent to the spectrum pricing and power control scheme given by the LEO satellite. It is noted that seamless coverage of the terrestrial server by LEO satellite is significant to ensure timely transmission of model parameters. Inter-satellite links are utilized to establish connections both within and between satellite constellations, enabling LEO satellites to relay data. Additionally, some of these satellites are equipped with onboard processing and storage capabilities, facilitating satellite-borne computing.

\subsection {Interference Model}

\begin{figure}[!t]
  \begin{center}
  \includegraphics[width=3.4in]{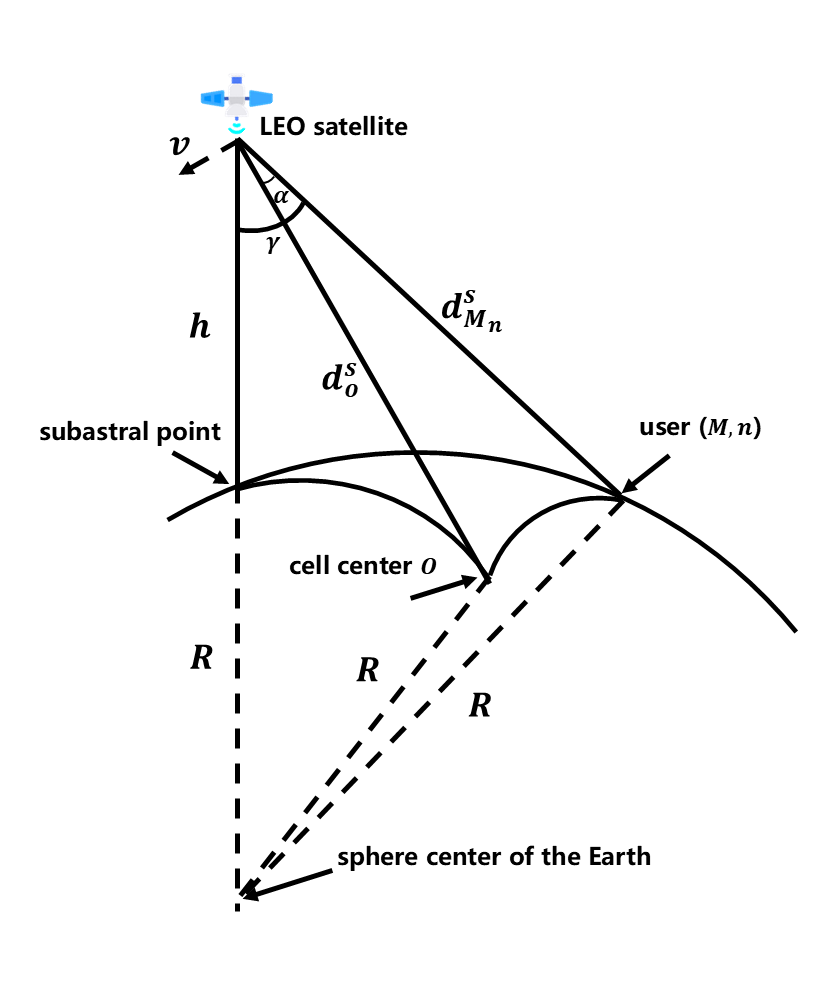}
  \caption{oblique projector}\label{sim_opt_eff_classFinv}
  \end{center}
\end{figure}

According to the satellite network architecture in this paper, the multi-beam antenna technique is applied. In this case, the Earth's surface is considered as a plane and the satellites project the beam onto the Earth's surface \cite{Li2023}. Unlike the propagation characteristics of high-orbiting satellites, there are more LEO satellites and more low Earth orbits, thus LEO satellites fly faster and cover a highly variable area. This makes the situation where most of the LEO satellites’ projection cells are not under the mode of orthographic projection even more prominent. Similar to the current existing work \cite{Li2019}, this paper considers the effect of the angle between the position of the selected user and the central line of the corresponding beam on the interference intensity. Thus, the angle $\alpha$ describing the deviation angle between user $(M, n)$ and cell center $o$ can be expressed as
\begin{equation}
\begin{split}
\begin{aligned}
    \alpha &= arccos(((d_o^s)^2 + (d_{Mn}^s)^2 - 2R^2(1-cos(d_{Mn}^o/R))) \\ 
    &\times (2d_o^sd_{Mn}^s)^{-1})
\end{aligned}
\end{split}
\end{equation}
where $d_o^s$ denotes the distance between the satellite and cell center, $d_{Mn}^s$ denotes the distance between satellite and user $(M, n)$, $R$ denotes the Earth radius, $d_{Mn}^o$ denotes the distance between cell center $o$ and user $(M, n)$ as shown in Fig. 2. 

Due to the high velocity of LEO satellites, the influence of Doppler shift \cite{dopplershift} on the spectrum quality can not be ignored. Frequency offsets may occur because of the relative motion between satellite and terrestrial users. Therefore, we consider the effect of Doppler shift, and the loss factor of LEO satellite $s$ can be expressed as
\begin{equation}
    \begin{split}
        F^d_s = \frac{1}{\frac{v_s}{C}cos(\gamma)\varpi + 1}
    \end{split}
\end{equation}
where $v_s$, $C$, $\gamma$ and $\varpi$ denote the relative velocity of LEO satellite $s$, the speed of light, the angle between the direction of motion and the direction of wave propagation and the fading coefficient. It is assumed that the idle spectrum of an LEO satellite is first divided into multiple channels by orthogonal frequency-division multiple access (OFDMA). Then each channel is leased to multiple users on the ground by time division multiple access. (TDMA). Inter-cell interference should be considered. For the uplink channel, the receiving power from user $n$ at cell $M$ can be expressed as 
\begin{equation}
    \begin{split}
        P = \frac{F^d_s\mathcal{P}_{n}g_{n}(\theta_n)G_{s}(\alpha_n^M)}{({\frac{4{\pi}d_{n}}{\lambda}})^2f_{n}(\theta_n)}
    \end{split}
\end{equation}
where $\mathcal{P}_{n}$ denotes the transmit power of satellite terminal, $\theta_n$ denotes the elevation angle from user $(M,n)$ to the satellite system, $g_{n}(\theta_n)$ denotes the antenna gain of user $(M,n)$ at the direction $\theta_n$, $\alpha_n^M$ is the derivation angle form user $(M,n)$ to the central line of cell $M$, $G_{s}(\alpha_n^M)$ is the satellite antenna gain of cell $M$ at the direction $\alpha_n^M$, $d_{n}$ is the straight-line distance between the user $(M,n)$ and the satellite system, $\lambda$ denotes the wavelength, $f_{n}(\theta_n)$ denotes the channel fading of user $(M, n)$ at the direction $\theta_n$.

The interference among the terrestrial cells can be given as 
\begin{equation}
    \begin{split}
        I = \sum_{H=1}^k\frac{F^d_s\mathcal{P}_{a}g_{a}(\theta_n)G_{s}(\theta_a^H)}{(4{\pi}d_{a}/\lambda)^2f_{a}(\theta_n)} \mu_a\rho_H^M
    \end{split}
\end{equation}
where $\mu_a$ denotes the active factor of user $a$ at cell $H$ which is related to the user’s service type. $\rho_H^M$ is the polarization isolation factor between cell $M$ and $H$.

Hence, the uplink Signal to Interference plus Noise Ratio (SINR) can be expressed as

\begin{equation}
    \begin{split}
        SINR = \frac{F^d_s\mathcal{P}_{n}g_{n}(\theta_n)G_{s}(\alpha_n^M)\lambda^2}{16{\pi}^2{d_{n}}^2f_{n}(\theta_n)I+\upsilon^2}
    \end{split}
\end{equation}
where $\upsilon^2$ denotes the power of noise. 

\subsection{Security Threats}

FL is introduced in this paper for machine learning collaborations among LEO satellite nodes. However, the system needs to rely on a trusted central server for global model aggregation. Besides, due to the potential misuse of spectrum by terrestrial users and malicious behaviors of the satellites in the FL, LEO satellite IoT is still facing system security and privacy preservation issues. The following threats are considered in the system.

1) \emph{Privacy Leakage and Global Model Tamping:} A central server is vulnerable to attack and may collude with other parties. 

2) \emph{Malicious Terrestrial Users:} A malicious terrestrial user may increase the uplink transmit power for a better QoS after leasing the spectrum.

3) \emph{Malicious Satellite Nodes:} A malicious satellite node may be fraudulent when transmitting the transaction data to the terrestrial server and may advertise fraudulent spectrum leasing services when they can not provide enough available spectrum.

In this paper, we propose a reputation-based blockchain combining FL to address the threats.

\subsection {Problem Formulation}

In LEO satellite IoT communication systems, satellites need to dynamically price spectrum based on the budgets of terrestrial users for spectrum resource leasing. This paper aims to maximize the benefits of LEO satellites while optimizing spectrum resource management. 

Typically, terrestrial users' budgets can refer to the QoS they receive. Thus, we formulate the budgets $\mathrm{B}_n$ of terrestrial user $n$ as
\begin{equation}
    \begin{split}
        \mathrm{B}_n = \zeta{SINR_n}
    \end{split}
\end{equation}
where $\zeta$ denotes the revenue coefficient. If the price of the idle spectrum set by the operator is below the budget, the terrestrial user may decide to lease the spectrum. We introduce an indicator function $\mathcal{I}_{n, s}[\mathrm{B}_n]$ to express the leasing intention of terrestrial users. Specifically, if the price set by the operator is lower than or equal to the budget $B_n$, the indicator would be $1$. Otherwise, the indicator would be $0$. Thus, the problem can be formulated as follows
\begin{align}
&\underset{\mathbf{s}}{\text{max}}\ \sum\limits_{n\in\mathbb{N}}\mathrm{P}_{s}\mathcal{I}_{n,s}[\mathrm{B}_n] \label{A}\\
&\textrm{s.t.} \quad {\mathrm{P}_s}\geq0,\ \forall s \in \mathbb{S}, \tag{\ref{A}{a}} \label{Aa}\\
&\quad\, \quad \kappa_{n}\in\mathbbm{1},\ \forall n \in \mathbb{N}, \tag{\ref{A}{b}} \label{Ab}\\
&\quad\, \quad \mathcal{I}_{n,s}[\mathrm{B}_n]\in\mathbbm{1},\ \forall n \in \mathbb{N}, \forall s \in \mathbb{S}, \tag{\ref{A}{c}} \label{Ac}\\
&\quad\, \quad {\mathcal{P}_{\text{min}}}\leq{\mathcal{P}_{n}}\leq{\mathcal{P}_{\text{max}}},\ \forall n \in \mathbb{N}, \tag{\ref{A}{d}} \label{Ad}\\
&\quad\, \quad {v_{\text{min}}}\leq{v_{s}}\leq{v_{\text{max}}},\ \forall s \in \mathbb{S}, \tag{\ref{A}{e}} \label{Ae}
\end{align}
where $\mathrm{P}_s$ denotes the price of the idle spectrum of satellite $s$, $\mathbb{S}$ denotes all the satellites in the network, $\mathbb{N}$ denotes all the users in the network, $\mathcal{P}_{\text{min}}$ and $\mathcal{P}_{\text{max}}$ denote the minimum transmit power and the maximum transmit power, $v_{\text{min}}$ and $v_{\text{max}}$ denote the minimum velocity and the maximum velocity of LEO satellite, $\kappa$ denotes the maximum number of idle channels that could be leased at the same time, which means terrestrial users could only lease zero or at most one channel at the same time. Eq. (7a) determines the range of the spectrum price set by LEO satellite operators. Eq. (7b) determines the maximum numb of idle channels that a user could lease at the same time. Eq. (7c) determines whether user $n$ decides to lease the idle spectrum based on the budget $\mathrm{B}_n$, Eq. (7d) and Eq. (7e) determine the range of transmit power and the velocity of LEO satellite.

Considering the above problem is not a naturally convex problem and involves many random variables, we use the DRL technique which is a model-free method. Model-free methods can quickly adapt to changes in the environment, such as fluctuating network conditions or varying interference levels. And as a widely used model-free method, DRL enables agents learn and adjust their strategies in real-time, making them suitable for dynamic communication networks. With sufficient model training, the DRL agents can effectively handle complex non-convex and sequential problems and provide near-optimal solutions.

\section{Framework of Privacy-aware Spectrum Pricing and Power Control}

\begin{figure}[!t]
  \begin{center}
  \includegraphics[width=3.5in]{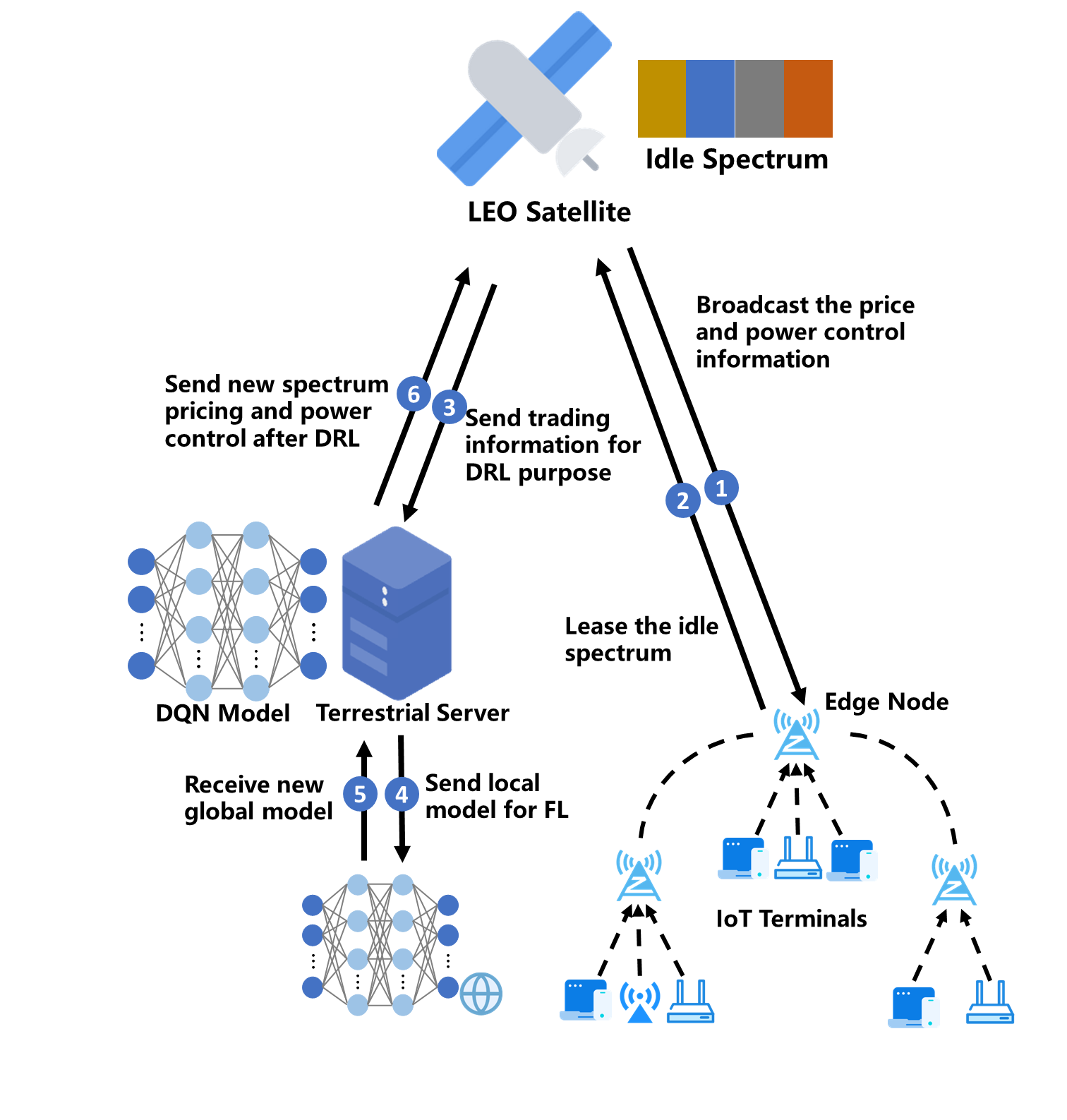}
  \caption{Operations in a local training round.}\label{sim_opt_eff_classFinv}
  \end{center}
\end{figure}

In this paper, we propose a privacy-aware spectrum pricing and power control scheme to facilitate spectrum resource management in the LEO satellite IoT. The scheme can be divided into three phases namely spectrum leasing and local training phase, blockchain-driven federated aggregation phase and global model release phase. The whole process is presented in Sec. IV-A. The modelings of the utility function and reinforcement learning environment are introduced in Sec. IV-B and Sec. IV-C respectively. And we present the details of blockchain-driven federated aggregation in Sec. IV-D.

\subsection{Whole Process}
The whole process of the scheme and the proposed framework is as follows.

\emph{Spectrum Leasing and Local Training Phase:} The operations in a local training round are presented in Fig. 3. Satellites are ready to lease their idle spectrum and set an initial price and initial uplink transmit power limit at first. Then, the satellite broadcasts the price and power control information to the terrestrial users (Label 1 in Fig. 3). After that, terrestrial users communicate with the satellites and decide to lease a certain spectrum (Label 2 in Fig. 3). After each round of trading, satellites transmit the collected trading information to their terrestrial servers for local DRL (Label 3 in Fig. 3) and update their reputation records based on terrestrial users' behaviors.  In addition, the servers record part of the trading information, verify the satellites' reputation records and send the local model to the Private Permissioned Blockchain network (Label 4 in Fig. 4) in preparation for the blockchain-driven federated aggregation in the next phase. After receiving the new global model (Label 5 in Fig. 3), servers transmit the new spectrum pricing and power control levels back to the satellite (Label 6 in Fig. 6).

\emph{Blockchain-driven Federated Aggregation Phase:} After several rounds of local training, each terrestrial server broadcasts the trading record in the Private Permissioned Blockchain network. The server with the highest reputation record aggregates the global model and performs the records package and block generation for the federated aggregation round. Then, the server gets a reward for the contribution and puts its reputation record to $0$.

\emph{Global Model Release Phase:} The server broadcasts the global model and each server starts the next round of training based on the global model.

\subsection{Modeling of Utility Function}

Throughout the process of leasing spectrum between terrestrial users and LEO satellites, LEO satellites price their idle spectrum and terrestrial users select spectrum to lease according to their required QoS. Specifically, to maximize their benefits, LEO satellites need to set the appropriate price for the spectrum. Whether a terrestrial user chooses to lease a certain spectrum and the amount of the satellite's revenue after leasing depends on the user's budget for spectrum leasing, the quality of the spectrum, the price of the spectrum, and the interference after leasing \cite{Lif2018}. Thus, the revenues of LEO satellites are closely related to the utilities of terrestrial users. In this case, the terrestrial users’ utility can be given as
\begin{equation}
    \begin{split}
    u_{n} = \mathrm{B}_n - \mathrm{P}_s
    \end{split}
\end{equation}
 We assume that if the required transmit power of a terrestrial user is bigger than the power control of the LEO satellite, then that user will not select this satellite’s spectrum to lease. For the revenue of the satellite, there are two components, one is the revenue of spectrum leasing and the other is the benefits obtained by contributing to the blockchain of the FL process which is mentioned in subsection D. And the utility at time slot $t$ can be given as
\begin{equation}
    \begin{split}
    u_{s} = \sum\limits_{n\in\mathbb{N}}\mathrm{P}_{s}\mathcal{I}_{n,s}[\mathrm{B}_n] + \mathcal{X}_n
    \end{split}
\end{equation}
where $\mathcal{X}_s$ denote the benefits obtained by contributing to the blockchain. 

\begin{figure}[!t]
  \begin{center}
  \includegraphics[width=3.5in, height=2.7in]{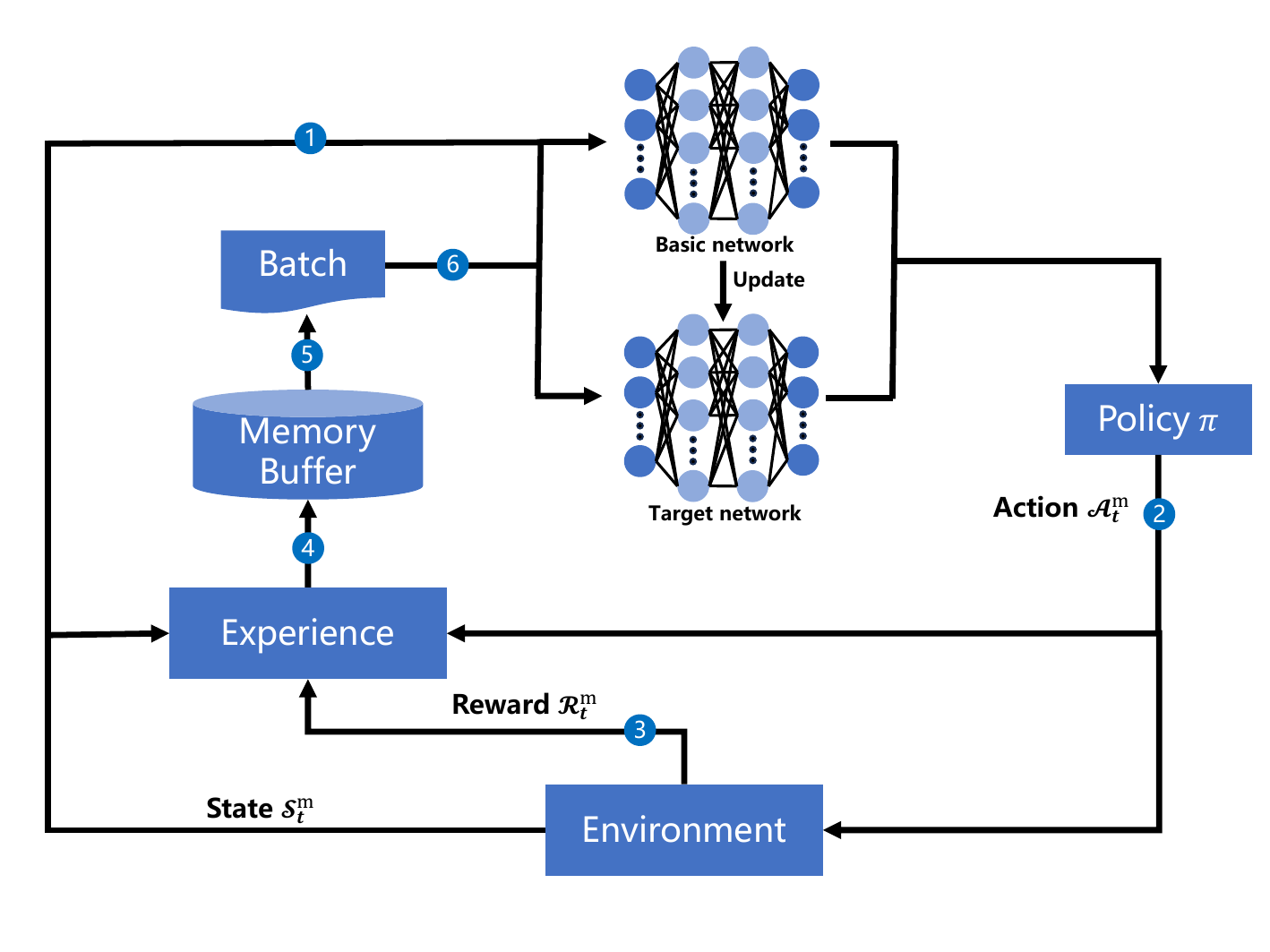}
  \caption{Framework of local DDQN.}\label{sim_opt_eff_classFinv}
  \end{center}
\end{figure}

\subsection{Modeling of Reinforcement Learning Environment}

In the process of spectrum leasing between LEO satellite IoT systems and terrestrial users, the price of spectrum is influenced by both the state of the satellites themselves and the conditions available to terrestrial users. Specifically, the idle spectrum’s status and uplink transmit power control status of the satellite, the terrestrial users' budgets for the leased spectrum and the interference after leasing the spectrum are all significant factors affecting the pricing of the satellite spectrum. However, such transaction information shared in the LEO satellite system is extremely limited due to the concerns for privacy preservation, which leads to unsatisfactory benefits of idle spectrum leasing at the satellite side, and large deviations in the QoS obtained by users at the terrestrial user side. For example, all the users in a cell do not have information about the number of users and their locations at the final leasing stage of the LEO satellite idle spectrum. This means that the final interference is also uncertain, which leads to variations in the QoS. Also, due to the uncertainty of the interference, terrestrial users tend to be conservative in their bids, making the satellite pricing of the idle spectrum lower than the benefit-maximizing price. Therefore, we introduce the DDQN, a model-free algorithm to find the optimal solution, where each LEO satellite performs as an agent. The framework of DDQN is illustrated in Fig. 4. 

\begin{algorithm}[!t]
\caption{DDQN-based Algorithm for Local Pricing and Power Controlling}
\label{alg:Framework}
\begin{algorithmic}[1]
    \STATE \textbf{Initialization:}
    \STATE \quad Basic network parameter $w$
    \STATE \quad Target network parameter $\hat{w}$
    \STATE \quad Learning rate $\delta$
    \STATE \quad Discount factor $\epsilon$
    \STATE \quad Target network parameter updating frequency $f$
    \FOR{each episode} 
    \STATE Initialize the state $\mathcal{S}_t^m$.
    \FOR{each step}
    \STATE Observe the current spectrum leasing and power control conditions from the environment.
    \STATE Select an action based on the target network and policy: select a random action $\mathcal{A}_t^m$ with probability $\vartheta$, select the action $\mathcal{A}_t^m=argmaxQ(\mathcal{S}_t^m, \mathcal{A}_t^m)$ with probability $1-\vartheta$.
    \STATE Execute action $\mathcal{A}_t^m$ to change the price of the spectrum or the control level of power.
    \STATE Receive a reward $\mathcal{R}_t^m$ and a new state $\mathcal{S}_{t+1}^m$.
    \STATE Store the experience $E=[\mathcal{S}^m_t, \mathcal{A}^m_t, \mathcal{S}^m_{t+1}, \mathcal{R}_t^m]$ to the memory buffer $M$.
    \STATE Draw randomly a mini-batch $\hat{M}$ from memory buffer $M$.
    \STATE Update the basic network parameter $w$.
    \IF{step mod $f$ == 0}
    \STATE Set target network parameter $\hat{w}$ equals to $w$.
    \ENDIF
    \ENDFOR
    \ENDFOR
\end{algorithmic}
\end{algorithm}

We first formulate the process of LEO satellite spectrum pricing as a Markov decision process (MDP) consisting of four parts: agent state space, action space, policy and reward function. Each agent continuously interacts with the environment while continuously changing its own policy to maximize reward. The specific details of the four elements are as follows.
\begin{itemize}
    \item \textbf{State:} The state of the agent can be described as
    \begin{equation}
        \mathcal{S}_s=[\mathrm{P}_S, {u'}_s]
    \end{equation}
    where ${u'}_s$ denotes the utility of satellite without considering the benefits from blockchain.
    \item \textbf{Action:} After obtaining the state, the satellite will choose an action $a_s$ to change the spectrum pricing to find a higher revenue. The action $\mathcal{A}_s$ of the agent can be described as 
    \begin{equation}
        \mathcal{A}_s=[\delta_s]
    \end{equation}
     where $\delta_s$ denotes the price level decision, and $\delta_s\in\{0,1\}$. $0$ means decrease one level, $1$ means increase one level.
    \item \textbf{Policy:} We define the policy $\pi(\mathcal{S}_{s, t+1} |\mathcal{S}_{t, s},\mathcal{A}_{s,t})$ to mapping from states to actions, which denote the probability that the agent $s$ selects action $\mathcal{A}_{s,t}$ from state $\mathcal{S}_{s, t}$ into a new state $\mathcal{S}_{s, t+1}$ at time slot $t$.
    \item \textbf{Reward:} To find an appropriate price to maximize the LEO satellites' revenue, the reward will play a key role in evaluating the learning policy. The reward in this paper can be given as
    \begin{equation}
        \mathcal{R}_{s, t}=
        \begin{cases}
            -1& {u'_{s, t} < u'_{s, t-1}}\\
            0& {u'_{s, t} = u'_{s, t-1}}\\
            1& {u'_{s, t} > u'_{s, t-1}}
        \end{cases}
    \end{equation}
    
\end{itemize}


To maximize the long-term cumulative reward, the agent needs to search for an optimal policy $\pi(\mathcal{S}_{s, t+1} |\mathcal{S}_{s, t},\mathcal{A}_{s,t})$ when interacting with the environment. During the process, agents execute an action $\mathcal{A}_{s,t}$, transitioning from the current state $\mathcal{S}_{s,t}$ to the next state $\mathcal{S}_{s,t+1}$. Specifically, each state transitioning of the agent is based on the transition probability. After each state transition, agents receive a reward $\mathcal{R}_{s,t}$ from the environment. The long-term accumulation reward is called the state-value function which is defined as 

\begin{equation}
    V^{\pi}(\mathcal{S})=\mathbb{E}_{\pi}\left[\sum\limits_{t=1}^{\infty}\gamma_t\mathcal{R}_{s,t}(\mathcal{S}_{s,t+1}, \mathcal{A}_{s,t})|\mathcal{S}_{s,t+1}=\mathcal{S}\right]
\end{equation}
Since the reward obtained after each interaction with the environment is immediate feedback, each decision is likely to have an impact on all subsequent states. Thus $\gamma_{s,t} \in (0,1]$ is a discount factor indicating the proportion of the future rewards' value of the current moment. And the optimal state-value function $V'(\mathcal{S})$ is defined as
\begin{equation}
    V'(\mathcal{S})={\mathop{max}\limits_\pi}V^\pi(\mathcal{S})
\end{equation}

In this paper, DDQN is introduced to address the MDP problems which can adapt to the environment with uncertainty. And the long-term accumulative reward is expressed by the Q-value function. Each agent has two neural networks which are basic network $\mathcal{B}$ and target network $\mathcal{T}$. The basic network $\mathcal{B}$ of each agent is updated in real-time while the target network $\mathcal{T}$ is updated based on the updating frequency factor $f$ to avoid overestimating the Q-value. The Q-value function can be given by 
\begin{equation}
    Q^\pi(\mathcal{S},\mathcal{A})=\mathbb{E}_{\pi}\left[\sum\limits_{t=1}^{\infty}\gamma_t\mathcal{R}_{s,t}(\mathcal{S}_{s,t},\mathcal{A}_{s,t})|\mathcal{S}_{t}=\mathcal{S},\mathcal{A}_{s,t}=\mathcal{A}\right]
\end{equation}
So the optimal Q-function is defined as
\begin{equation}
    Q'(\mathcal{S},\mathcal{A})={\mathop{max}\limits_\pi}Q^\pi(\mathcal{S},\mathcal{A})
\end{equation}
And based on the Bellman Optimality Equation, the Q-value function can be defined as 
\begin{equation}
\begin{aligned}
    Q&(\mathcal{S}_{s,t},\mathcal{A}_{s,t})=\\
    &\mathcal{R}_{s,t}+\gamma_tQ(\mathcal{S}_{s,t+1},argmax_{\mathcal{A}_{s,t}\in\mathcal{A}}Q(\mathcal{S}_{s,t+1},\mathcal{A}_{s,t};\mathcal{B}_{s,t});\mathcal{T}_{s,t})
\end{aligned}
\end{equation}
Then the Q-value function is updated by
\begin{equation}
\begin{aligned}
    Q_{t+1}&(\mathcal{S}_{s,t},\mathcal{A}_{s,t})=(1-l)Q_t(\mathcal{S}_{s,t},\mathcal{A}_{s,t})+l(\mathcal{R}_{s,t}+\\
    &{\gamma_t}Q(\mathcal{S}_{s,t+1},argmax_{\mathcal{A}_{s,t}\in\mathcal{A}}Q(\mathcal{S}_{s,t+1},\mathcal{A}_{s,t};\mathcal{B}_{s,t});\mathcal{T}_{s,t}))
\end{aligned}
\end{equation}
where $l\in(0,1]$ denotes the learning rate.

The user selects an action to execute based on $\vartheta$-policy in each training step, which can be expressed as
\begin{equation}
    \mathcal{A}_{s,t+1} = 
    \begin{cases}
    \mathcal{A}_{random}& {P=\vartheta}\\
    argmax_{\mathcal{A}_{s,t+1}\in\mathcal{A}}Q(\mathcal{S}_{s,t+1},\mathcal{A}_{s,t+1})& {P=1-\vartheta}
    \end{cases}
\end{equation}

The details and the whole process of the local DDQN model training are presented in Fig. 4 and Algorithm 1. Each LEO satellite acts as an agent and first initializes its basic network parameter $w$ and target network parameter $\hat{w}$ (Line 2-3 in Algorithm 1). And they perform $E$ episodes in each local training round. When performing local DDQN, the LEO satellite first observes the current state $\mathcal{S}_{s,t}$ which is the current spectrum leasing and power control conditions (Label 1 in Fig. 4, Line 10), and selects an action $\mathcal{A}_{s,t}$ based on the target network and the policy $\pi$ (Label 2, Line 11). The agent randomly selects an action from action space $\mathcal{A}$ with probability $\vartheta$ and selects the action with maximum Q-value with probability $1-\vartheta$. After executing the action $\mathcal{A}_{s,t}$, an immediate reward $\mathcal{R}_{s,t}$ and a new state $\mathcal{S}_{s,t+1}$ are obtained (Label 3, Line 13) which construct the experience together with the action $\mathcal{A}_{s,t}$ and state $\mathcal{S}_{s,t}$. Then the experience of that episode is stored in the memory buffer (Label 4, line 14). Next, a batch is randomly drawn from the memory buffer for updating the basic network (Labels 5-6, Line 15). The target network will be updated for every $f$ local training round.

\subsection{Blockchain-driven Federated Aggregation}

\begin{figure}[!t]
  \begin{center}
  \includegraphics[width=3.5in, height=2.8in]{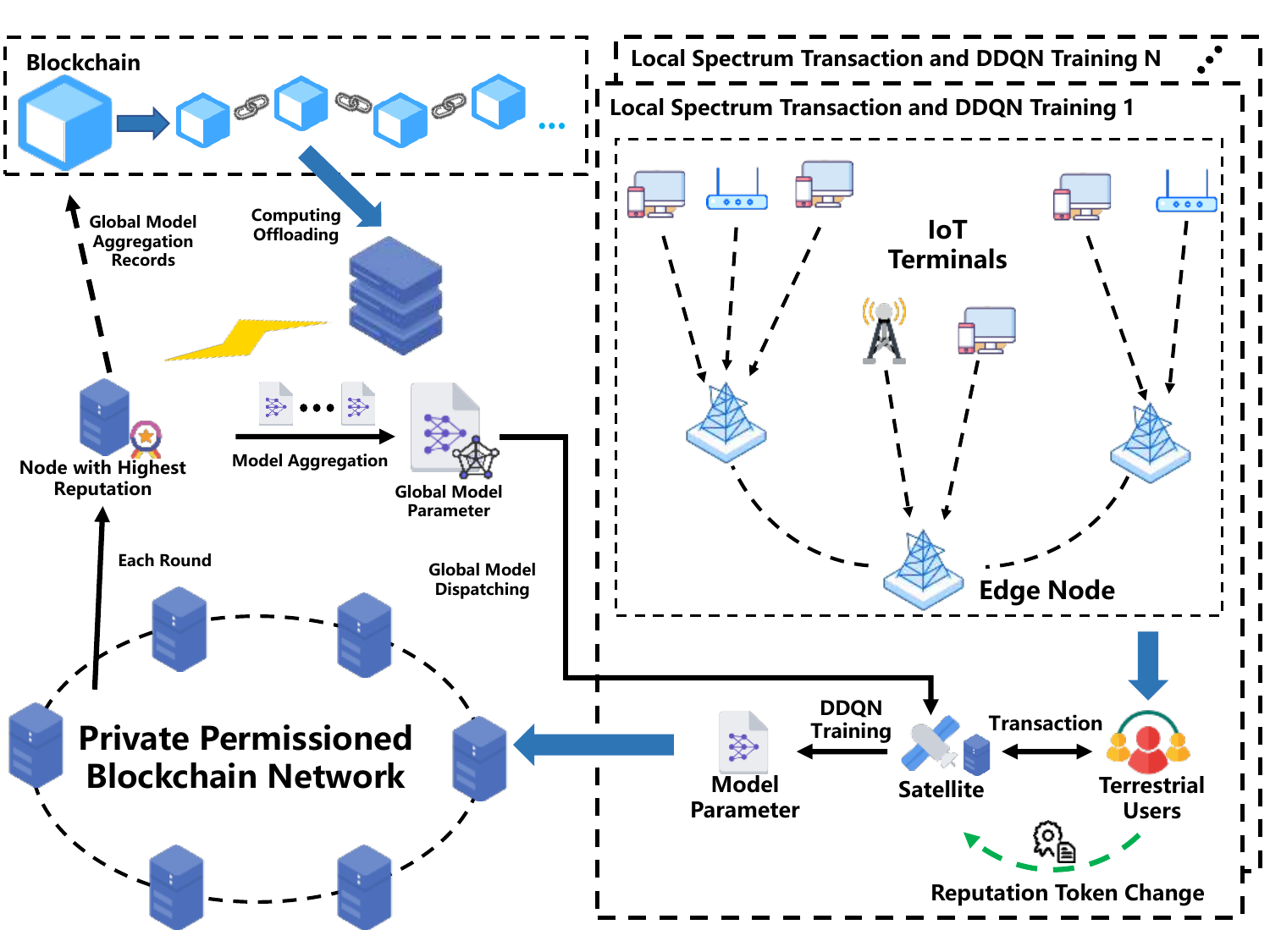}
  \caption{Framework of Reputation-based Blockchain Network and FL Process}\label{sim_opt_eff_classFinv}
  \end{center}
\end{figure}

In our work, the blockchain is deployed among terrestrial servers of LEO satellites. This deployment not only drives the operators of LEO satellites to supervise and optimize the power control for those terrestrial users who lease the spectrum for communication services but also guarantees the auditability of the global model aggregation in federated learning. 

First, in each global model aggregation round, an agent of LEO satellite will be chose based on a reputation consensus mechanism to aggregate the global model, which will receive a reward for the contribution to the blockchain. The agent for generating the blockchain is decided by a reputation record. To maintain a good reputation record, the operators of LEO satellites need to supervise and adjust the power control level so that no terrestrial users conduct malicious behaviors such as exceeding the transmit power maliciously. 

Secondly, to improve the model training efficiency while protecting sensitive information, FL is introduced to the satellite IoT in this paper. In the satellite IoT, data owned by each LEO satellite can not be shared to better improve the efficiency of dynamic spectrum pricing and trading because of the sensitivity of the transaction information involved. Instead of obtaining the original sensitive data, FL aggregates the local training model parameters of each LEO satellite to form a global model and sends it back, which is an effective improvement in the related issue. However, in traditional FL, centralized global model aggregation remains a threat to the privacy-preserving of local devices. If the server aggregating the global model is malicious or the server is attacked, the spectrum pricing and spectrum trading of the whole IoT system will be paralyzed. Thus, based on the feature of LEO satellite IoT communicating and transacting with terrestrial users, the deployment of the reputation-based blockchain makes sure that the aggregation of global model is not centralized and the parameter of global model in each global training round is traceable after each local training in LEO satellite, thus enhancing the suitability of global model aggregation.

As shown in Fig. 5, the network of Reputation-based Blockchain involves several components.
\begin{itemize}
    \item \textbf{Satellite:} Satellite provides the idle spectrum to terrestrial users and trains the local DDQN model to search the optimal spectrum price and power control.
    \item \textbf{Terrestrial users:} Terrestrial users are the spectrum demanders. They decide whether to lease the idle spectrum provided by a certain satellite based on their budget and requirements for spectrum quality.
    \item \textbf{Reputation token:}  Each satellite that is a member of a private permissioned blockchain network has a reputation token record. The reputation token record of each satellite changes for each round of dynamic spectrum access by terrestrial users. If there are no users with malicious behavior in this access round, the number of reputation tokens for that satellite is the original number of reputation tokens plus the newly acquired reputation tokens. Instead, if there is a malicious user in this access round, the number of reputation tokens for the satellite is the original number of reputation tokens minus the penalty incurred for the malicious user. The reputation record of a satellite can be expressed as
    \begin{equation}
    {Rep}_{s,t} = 
    \begin{cases}
    {Rep}_{s,t-1}+\mathcal{V}_t^{acc}& {no\;malicious\;users}\\
    {Rep}_{s,t-1}-\mathcal{V}_t^{mal}& {malicious\;users\;appear}
    \end{cases}
\end{equation}
    where $Rep_t^m$ denotes the reputation token, $\mathcal{V}_t^{acc}$ denotes the newly acquired reputation token based on the situation that no malicious users appear and $\mathcal{V}_t^{mal}$ denotes the newly lost reputation token based on the situation that malicious users appear respectively. Then $\mathcal{V}_t^{acc}$ and $\mathcal{V}_t^{mal}$ can be expressed as
    \begin{equation}
    \mathcal{V}_t^{acc} = \hat{p}l_{s,t}N^{acc}_{pow,t}
    \end{equation}
    \begin{equation}
    \mathcal{V}_t^{mal} = \hat{p}l_{s,t}N^{mal}_{pow,t}
    \end{equation}
    where $\hat{p}$ denotes the reputation coefficient, $l_{s,t}$ denotes the power control level of the satellite, $N^{acc}_{pow,t}$ denotes the number of normal users whose transmit power is lower than $l_{s,t}$, $N^{mal}_{pow,t}$ denotes the number of malicious users. 
    The malicious behavior of ground users is as follows. 1) The number of terrestrial users accessing the spectrum exceeds the limit. 2) Terrestrial users accessing the satellite's spectrum without meeting the required power level. 3) Terrestrial users access the spectrum for too long or too short a period of time based on the spectrum lease contract.
    \item \textbf{Edge node:} {Edge nodes are responsible for verifying transaction users, conducting spectrum transactions and saving transaction records. Each terrestrial user pays money to the satellite through the edge node.}
    \item {\textbf{Satellite terrestrial server:} 
    LEO satellites are generally compact with limited computational resources, thus a satellite terrestrial server is required to take up most of the computational procedures. LEO satellites have limited storage and finite bandwidth, which can delay data transmission without the involvement of satellite terrestrial server. In terms of the communication latency as well, the satellite terrestrial server can help to maintain a stable link for continuous spectrum management despite of the frequently dropped connections due to the nature of LEO satellite which moves quickly relative to fixed ground stations. In details, satellite terrestrial servers are responsible for the data computing and model training of their corresponding satellites. Besides, these satellite terrestrial servers participate in the private permissioned Blockchain network operation to make the data auditable.}
\end{itemize}

\begin{algorithm}[!t]
\caption{Process of Global Model Aggregation and Blockchain Updating}
\label{alg:Framework}
\begin{algorithmic}[1]
    \STATE \textbf{Initialization:}
    \STATE \quad Model Aggregation frequency $f'$
    \FOR{each global training round}
    \STATE Each node updates the reputation $Rep_t^m$ based on the reputation mechanism.
    \IF{global training round mod $f'$ == $0$}
    \STATE Each node broadcasts the model parameter to each of the other nodes of the private permissioned blockchain network.
    \STATE Get the list by ranking the reputation of each node from highest to lowest.
    \FOR{each node of the list}
    \IF{node is online}
    \STATE The node aggregates the global model and generates the block.
    \STATE The node sends the model to other nodes who locally train the model.
    \STATE The node receives the reward for model aggregation and block generation.
    \STATE End this round of global training.
    \ELSIF{node is offline}
    \STATE Select the next node.
    \ENDIF
    \ENDFOR
    \ENDIF
    \ENDFOR
\end{algorithmic}
\end{algorithm}

Algorithm 2 shows the process of global model aggregation and blockchain updating. Model aggregation frequency $f'$ is initialized first. After each round of spectrum transactions between satellites and the covered area's terrestrial users, each LEO satellite filters the received information from the transactions and then send the the information to the corresponding satellite terrestrial server. Next, each satellite terrestrial server in the private permissioned blockchain network updates the reputation record of its LEO satellite based on the terrestrial users' behaviors. When the global training round mod $f'$ equals $0$, it is the round for global aggregation and updating. Each satellite terrestrial server broadcasts its local parameter to each of the other satellite terrestrial servers of the blockchain network. Then each LEO satellite's reputation is ranked from highest to lowest. The aggregation priority of each satellite terrestrial server is represented in list order. If the server is online, then it becomes the aggregation node in this round. And if the server is offline, the next server will be considered. For the aggregation server, once the server is confirmed, it first aggregates the global model and generates the block, and then sends the model to other satellite terrestrial servers who locally train the model. After that, the operator of the LEO satellite will receive the reward for the contribution. The inter-satellite links can improve energy efficiency by reducing the need for satellites to continuously establish high-power downlinks to ground stations, which is especially energy-intensive for LEO satellites. By using lower-power ISLs for the majority of communications, satellites can conserve energy, prolonging operational lifetime while still supporting frequent model updates.

\begin{figure}[t!]
    \begin{subfigure}[t]{1\linewidth}
           \centering
           \includegraphics[width=3.5in]{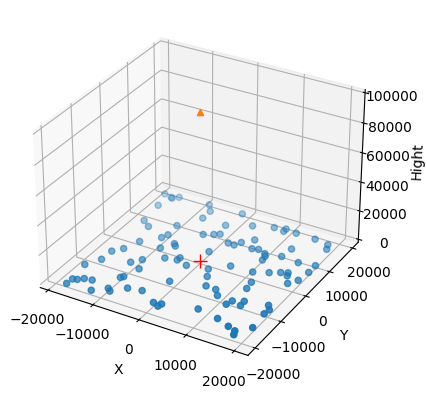}
            \caption{}
            \label{fig:Case1}
    \end{subfigure}
    
    \begin{subfigure}[t]{1\linewidth}
            \centering
            \includegraphics[width=3.5in]{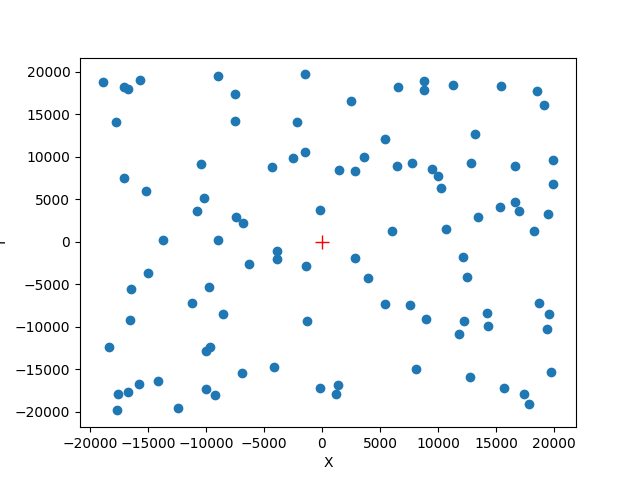}
            \caption{}
            \label{fig:b}
    \end{subfigure}
    \caption{Distribution of terrestrial users and LEO satellite}
\end{figure}

\begin{figure*}[t!]
    \begin{subfigure}[t]{0.31\linewidth}
           \centering
           \includegraphics[width=2.35in]{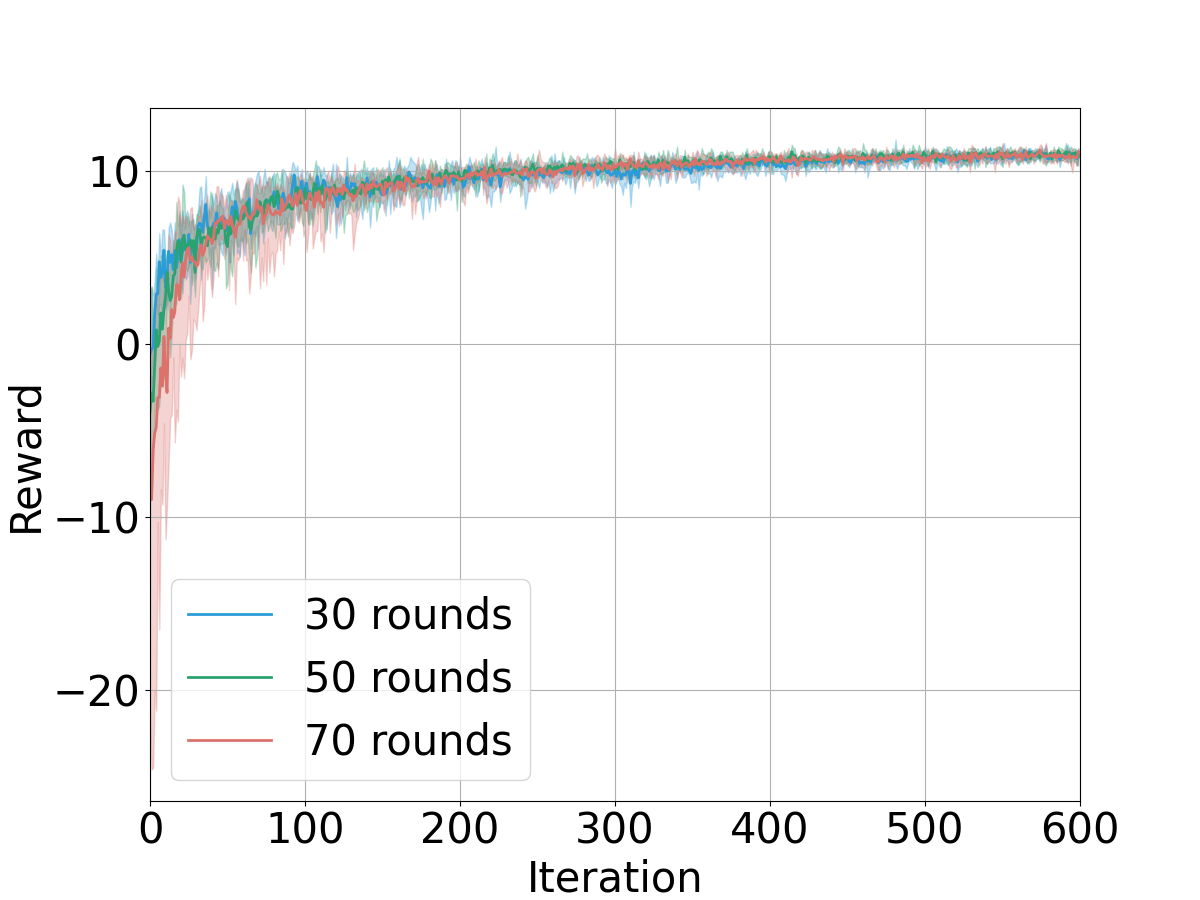}
            \caption{10 Agents}
            \label{fig:Case1}
    \end{subfigure}
    \begin{subfigure}[t]{0.31\linewidth}
            \centering
            \includegraphics[width=2.35in]{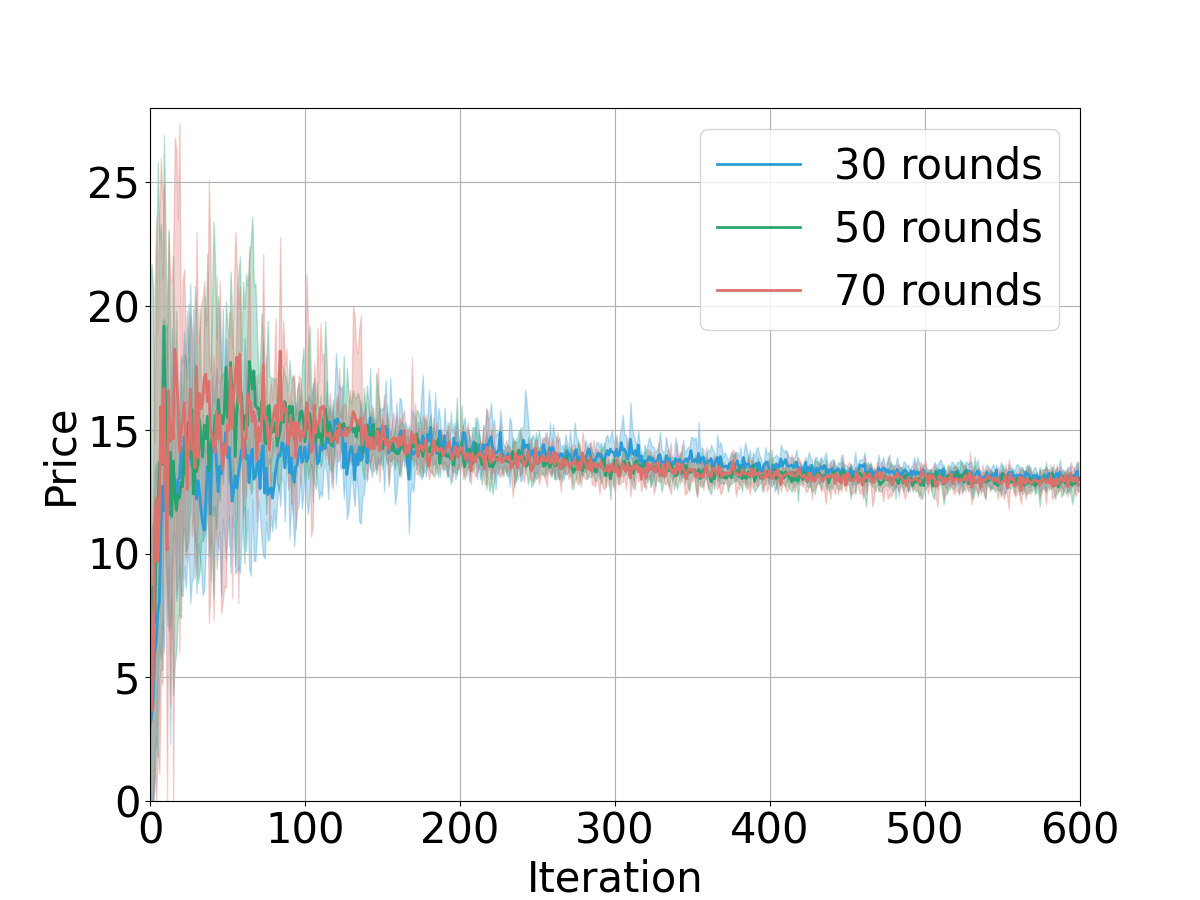}
            \caption{10 Agents}
            \label{fig:b}
    \end{subfigure}
    \begin{subfigure}[t]{0.31\linewidth}
            \centering
            \includegraphics[width=2.35in]{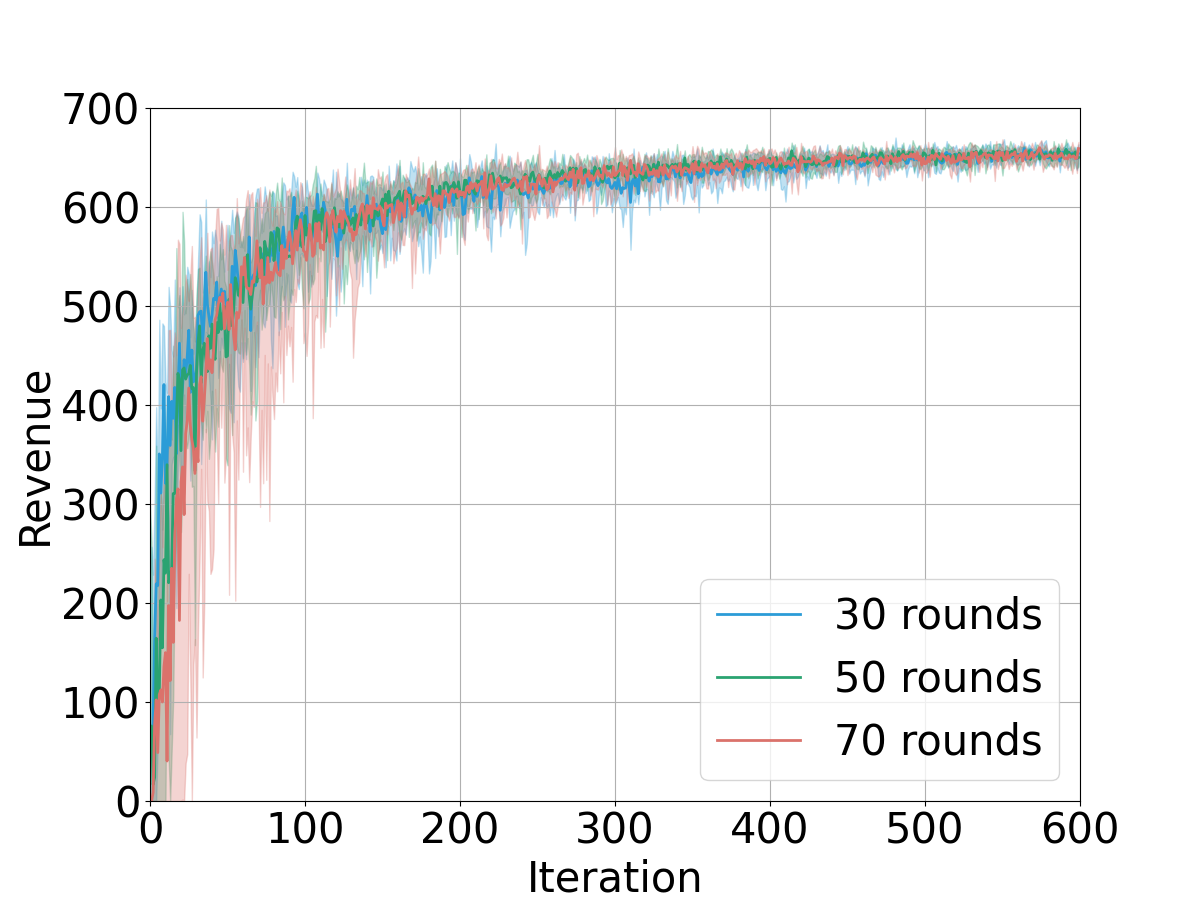}
            \caption{10 Agents}
            \label{fig:b}
    \end{subfigure}

    \begin{subfigure}[t]{0.31\linewidth}
           \centering
           \includegraphics[width=2.35in]{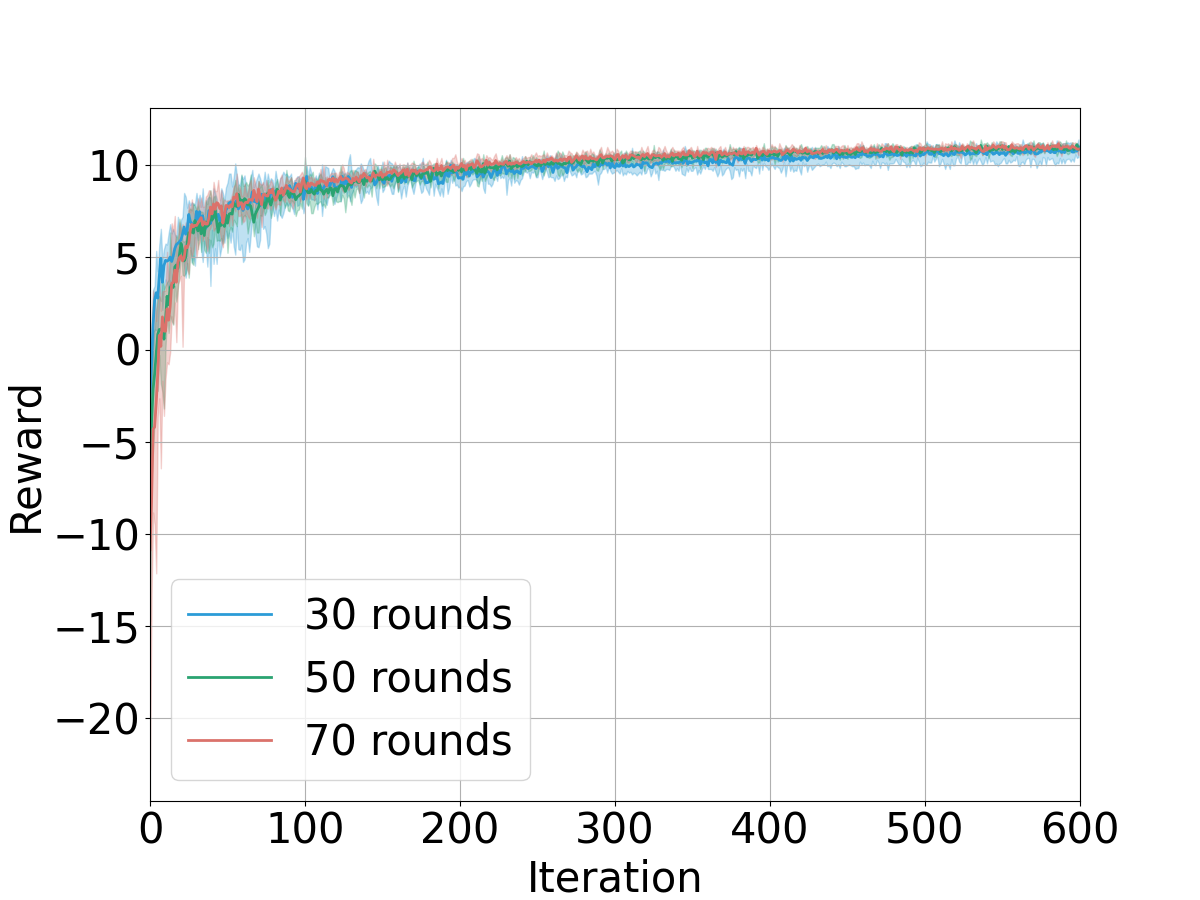}
            \caption{20 Agents}
            \label{fig:Case1}
    \end{subfigure}
    \begin{subfigure}[t]{0.31\linewidth}
            \centering
            \includegraphics[width=2.35in]{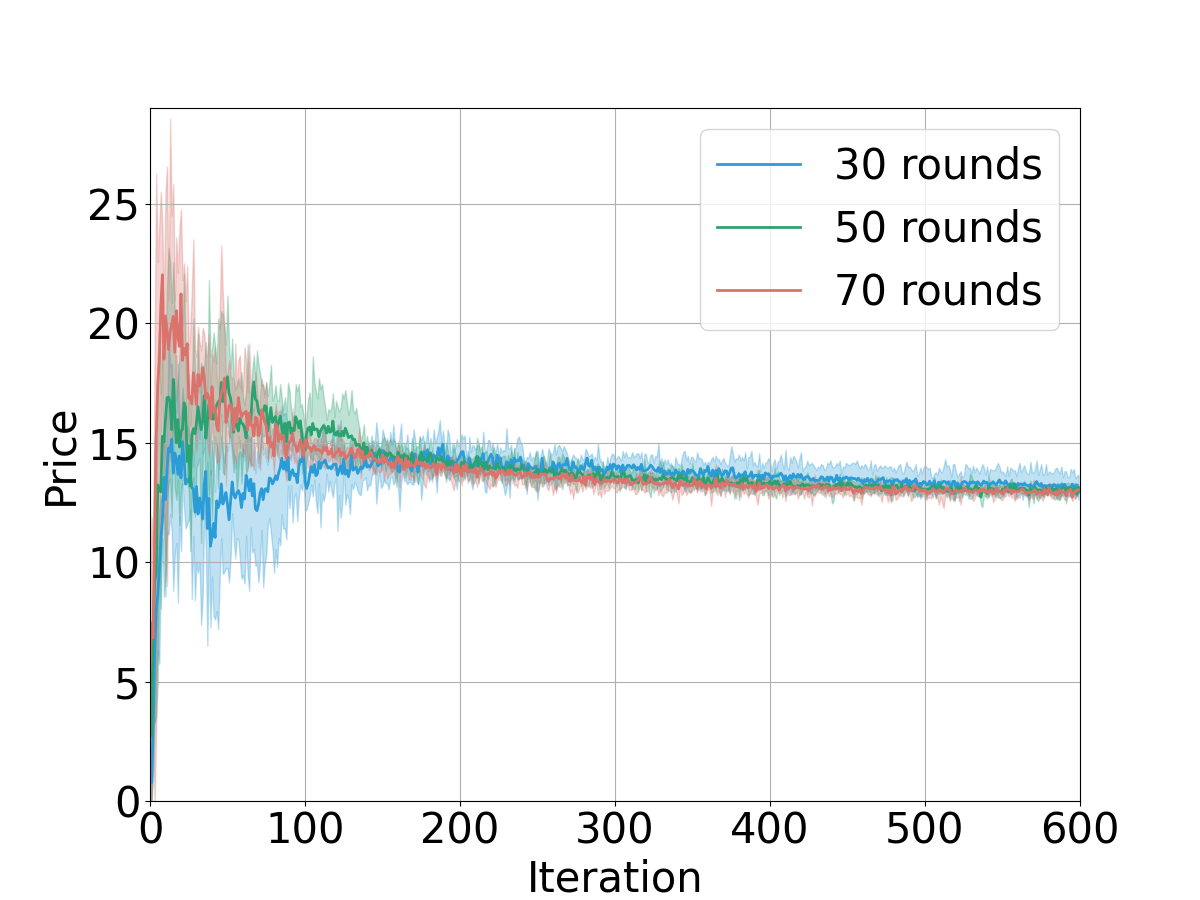}
            \caption{20 Agents}
            \label{fig:b}
    \end{subfigure}
    \begin{subfigure}[t]{0.31\linewidth}
            \centering
            \includegraphics[width=2.35in]{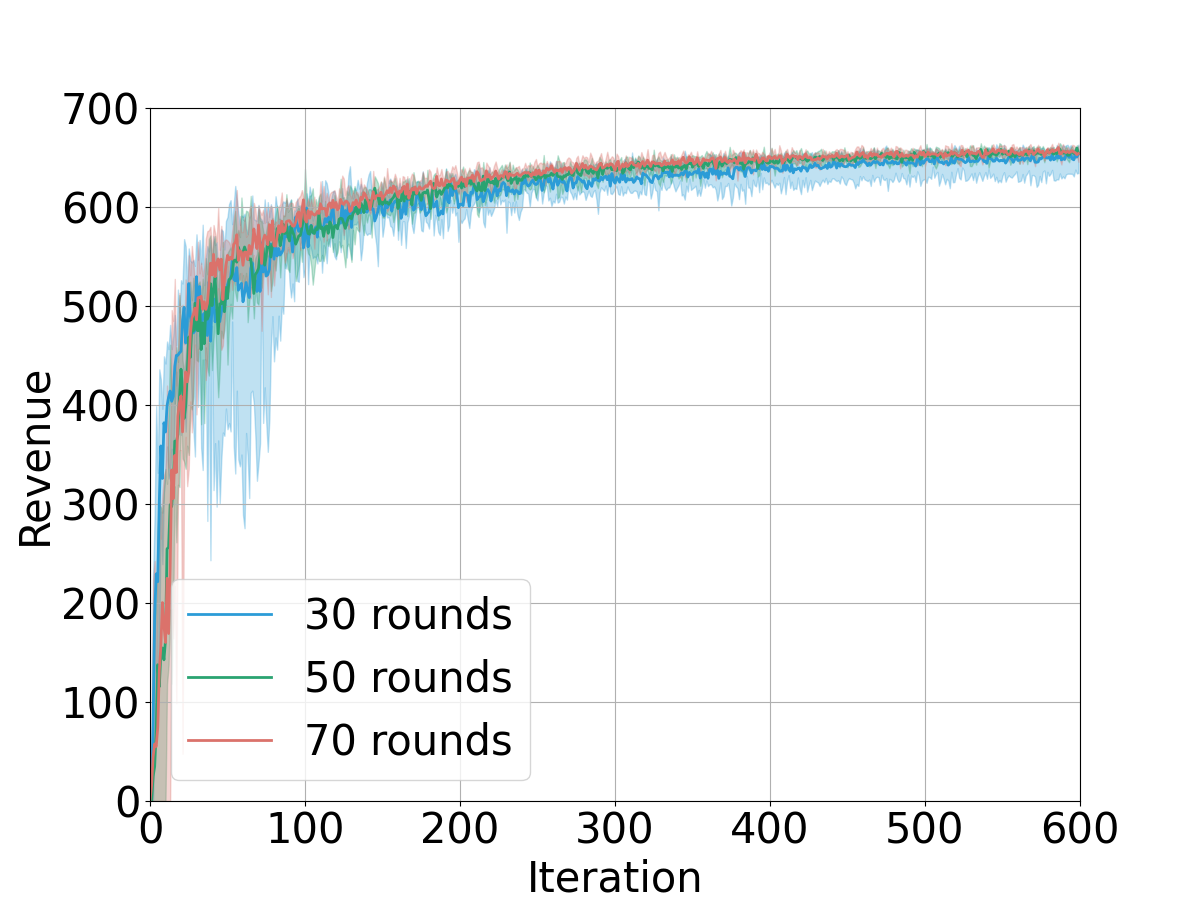}
            \caption{20 Agents}
            \label{fig:b}
    \end{subfigure}

    \begin{subfigure}[t]{0.31\linewidth}
           \centering
           \includegraphics[width=2.35in]{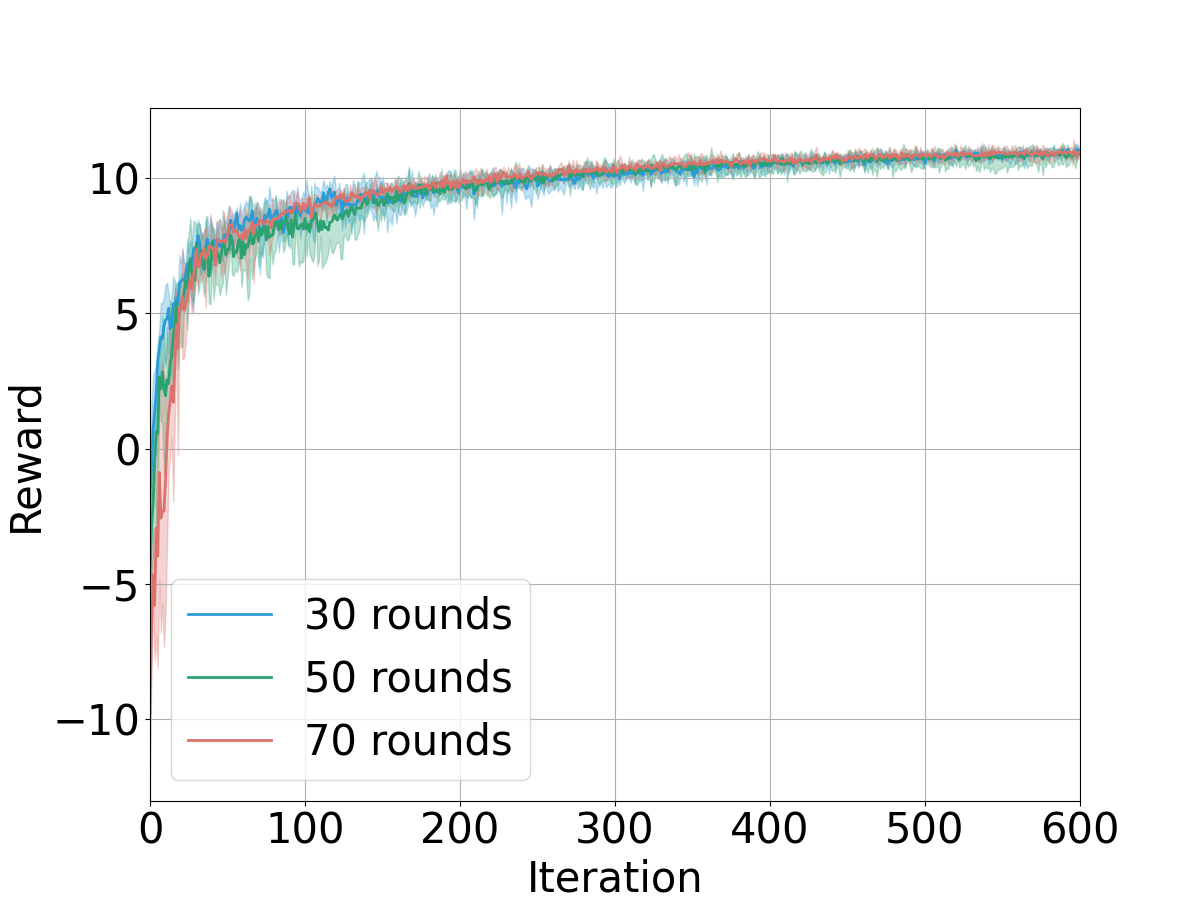}
            \caption{30 Users}
            \label{fig:Case1}
    \end{subfigure}
    \begin{subfigure}[t]{0.31\linewidth}
            \centering
            \includegraphics[width=2.35in]{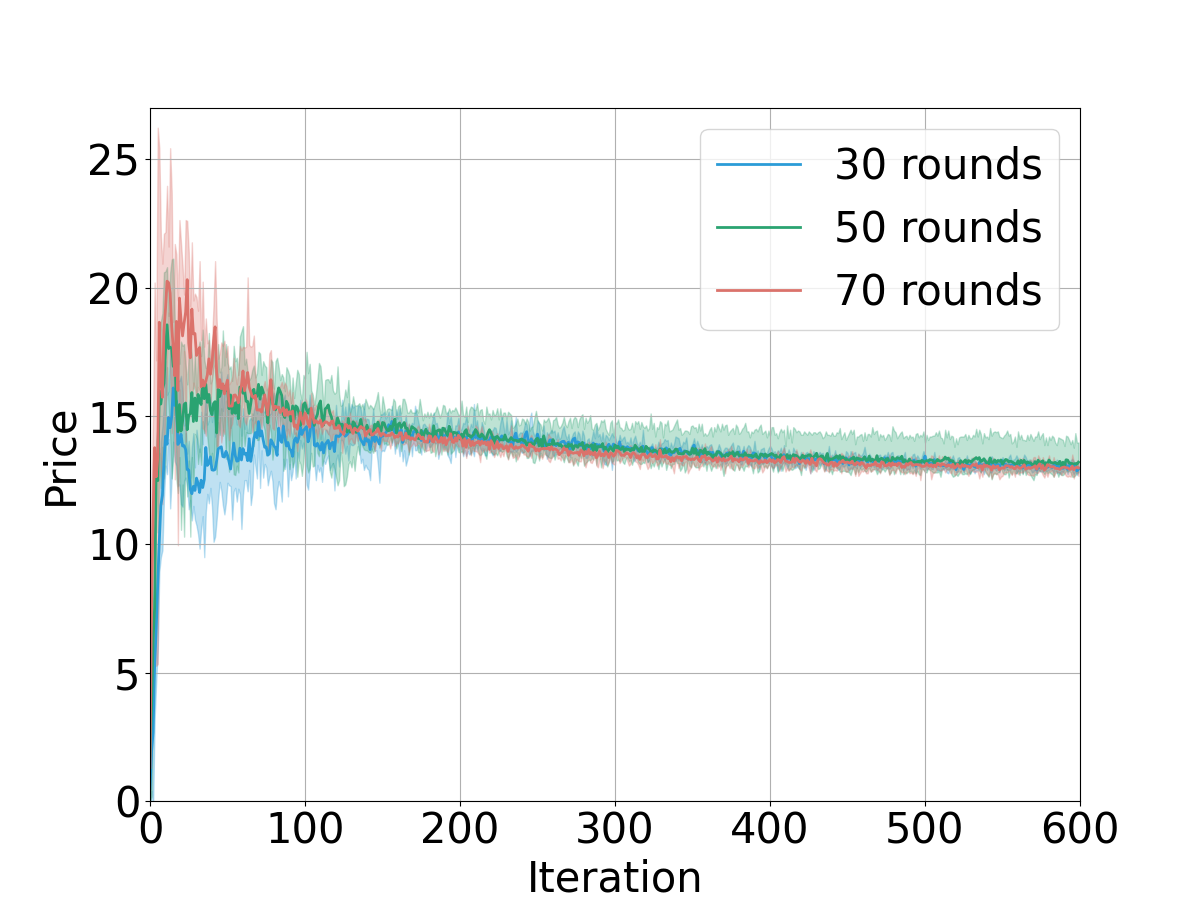}
            \caption{30 Agents}
            \label{fig:b}
    \end{subfigure}
    \begin{subfigure}[t]{0.31\linewidth}
            \centering
            \includegraphics[width=2.35in]{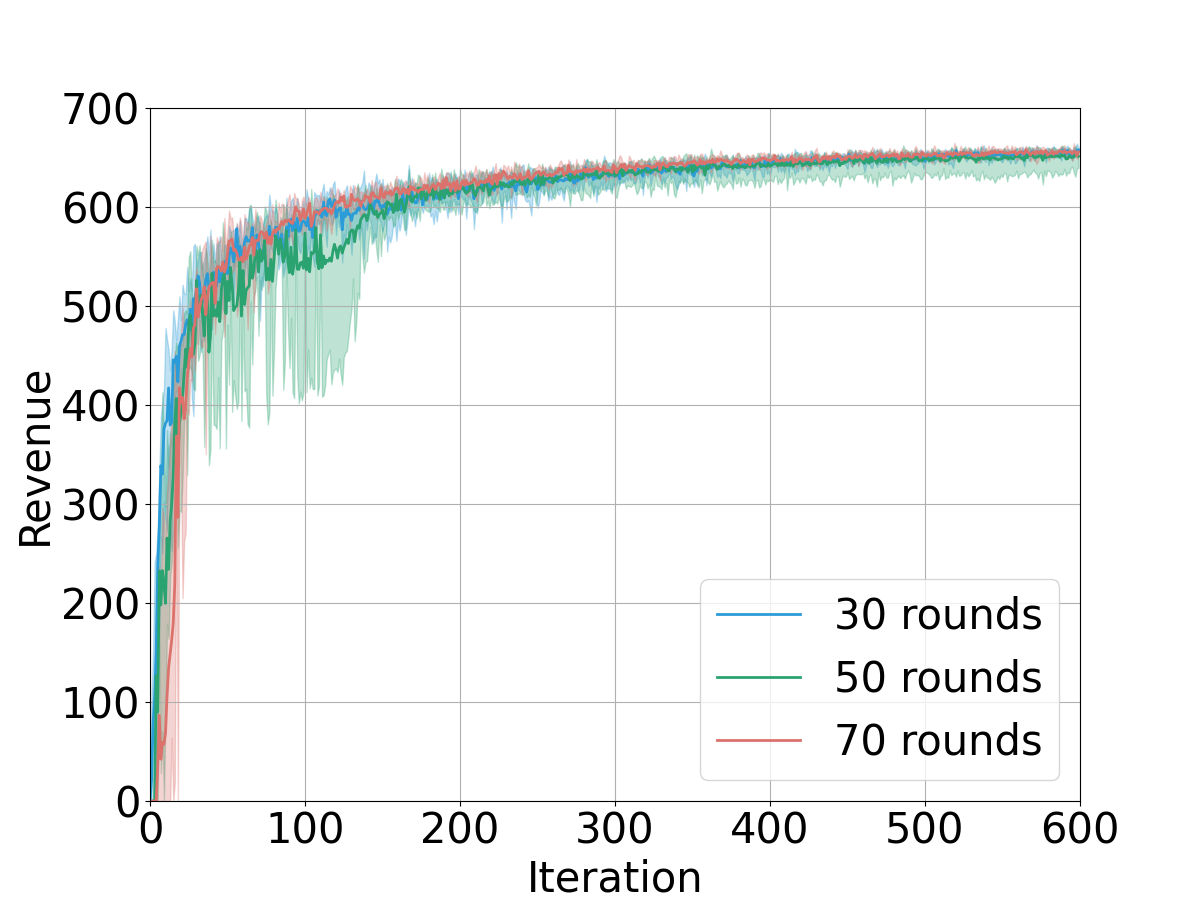}
            \caption{30 Agents}
            \label{fig:b}
    \end{subfigure}
    \caption{Average reward, price and revenue with different numbers of FL agents.}
\end{figure*}

\subsection{Complexity and Computational Analysis}
Let $\mathcal{N}$, $\mathcal{L}_i$ define the training layer and the number of the neurons in the $i$-th layer. Thus, computational complexity in each training time step for each agent is $O(\sum_{i=0}^\mathcal{N}\mathcal{L}_i\mathcal{L}_{i+1})$. And let $\mathcal{M}$, $\mathcal{K}$ and $\mathcal{E}$ denote the number of the trained models, total training round and episodes in each training round, respectively. The computational complexity can be expressed by $O(\mathcal{M}\mathcal{K}\mathcal{E}\sum_{i=0}^\mathcal{N}\mathcal{L}_i\mathcal{L}_{i+1})$ \cite{Jiang2020, Yang2022, Li2021}.

For the local DDQN training phase for each LEO satellite, the computational complexity of each LEO satellite can be expressed by $O(\mathcal{K}\mathcal{E}\sum_{i=0}^\mathcal{N}\mathcal{L}_i\mathcal{L}_{i+1})$. It is noted that local training is parallelized across satellites, so increasing the number of satellites primarily increases total distributed computational complexity rather than slowing any agent's training. The high local training workload can be performed offline for a finite number of episodes on terrestrial servers to avoid straining the satellite, which guarantees the feasibility of the training process by leveraging terrestrial computing resources even as the network scales up. 


\begin{table*}
    \caption{Comparisons for complexity of Aggregator selection, blockchain verification, aggregation, and block broadcast in different consensus mechanisms.}
    
    \label{tab:Notations}
    \centering
    \begin{tabular}{|c|cccc|}
    \hline
         Mechanism & Aggregator Selection & Blockchain Verification & Aggregation & Block Broadcast \\
         \hhline{|=|====|}
         Reputation-based & $O(\mathcal{Y}log\mathcal{Y})$ & $O(\mathcal{Y}||\hat{w}||)$ & $O(1/\sqrt{\sum_{t=1}^{T^{FL}}\mathcal{D}(t)})$ & $O(\mathcal{Y}||\hat{w}||)$\\
         \hline
         PoW & $O(2^{\varsigma})$ & $O(1)$ & $O(1/\sqrt{\sum_{t=1}^{T^{FL}}\mathcal{D}(t)})$ & $O(\mathcal{Q}||\hat{w}||)$\\ 
         \hline
         PoS & $O(\mathcal{Q}log\mathcal{Q})$ & $O(\mathcal{Q}||\hat{w}||)$ & $O(1/\sqrt{\sum_{t=1}^{T^{FL}}\mathcal{D}(t)})$ & $O(\mathcal{Q}||\hat{w}||)$ \\
         \hline
         BFT & $O(1)$ &  $O(\mathcal{Q}^2||\hat{w}||)$ & $O(1/\sqrt{\sum_{t=1}^{T^{FL}}\mathcal{D}(t)})$ & $O(\mathcal{Q}^2||\hat{w}||)$\\
         \hline
    \end{tabular}
\end{table*}

For the FL phase, the complexity consists of four parts: aggregator selection, blockchain verification, aggregation and Block Broadcast. Let $\mathcal{Q}$, $\mathcal{Y}$, $||\hat{w}||$ denotes the total number of LEO satellites, the number of participating LEO satellites in the aggregation blockchain network and the size of block. Thus, we can find the aggregator with the highest reputation by scanning through all reputation scores in $O(\mathcal{Y}log\mathcal{Y})$. The computational complexity of blockchain verification can be expressed as $O(\mathcal{Y}||\hat{w}||)$. The communication overhead can be expressed as $O(\mathcal{Y}^2)$. Let $\mathcal{D}(t)$ denote the number of devices involved at time slot $t$. Hence, the computational complexity for the global model aggregation is $O(1/\sqrt{\sum_{t=1}^{T^{FL}}\mathcal{D}(t)})$. Table III compares the computational complexity of the proposed reputational-based consensus mechanism with Proof of Work (PoW), Proof of Stake (PoS) and Byzantine Fault Tolerance (BFT), where $\varsigma$ denotes mining difficulty. It is noted that the number of the participating nodes is much less than the total number nodes, which leads to faster consensus finalization in large scale LEO satellites IoT scenarios. Besides, there is no need for mining in the proposed reputation-based consensus mechanism compared to PoW, resulting in lower energy consumption. Although the complexity for the aggregator selection of the proposed scheme is more than BFT, the complexity for blockchain verification and block broadcast is much less than the BFT due to the less validators and communication rounds. In a large-scale deployment, reliable and frequent communication is required for coordinating learning across satellites. Each round of federated learning involves satellites and their corresponding terrestrial server, transmitting their local model updates and receiving the aggregated global model. This overhead grows roughly linearly with the number of satellites. The proposed design mitigates this by leveraging inter-satellite links rather than routing everything through terrestrial stations, which allows satellites to communicate updates with lower power and latency, saving energy by avoiding continuous high-power downlink.

\emph{Lemma 1:} The formulated problem Eq. (7) is a non-convex problem.

\emph{Proof:} The indicator function $\mathcal{I}_{n, s}[\mathrm{B}_n]$ is discontinuous and non-differentiable at the boundary where $\mathrm{P}_s = \mathrm{B}_n$, violating the criteria for convexity.

\emph{Lemma 2:} $SINR(v_s) = \frac{\mathcal{P}_{n}g_{n}(\theta_n)G_{s}(\alpha_n^M)\lambda^2}{(16{\pi}^2{d_{n}}^2f_{n}(\theta_n)I+\upsilon^2)(\frac{v_s}{C}cos(\gamma)\varpi + 1)}$ is monotonically decreasing with the increasing of relative velocity $v_s$ of LEO satellite.

\emph{Proof:} The first order derivatives of $SINR(v_s)$ is derived as 
\begin{equation}
    \begin{split}
          SINR'(v_s) = -\frac{cos(\gamma)\varpi}{C(\frac{v_scos(\gamma)\varpi}{C}+1)^2}<0
    \end{split}
\end{equation}
Therefore, according to Eq. (23), $SINR(v_s)$ is a decreasing function.

The satellite’s main beam footprint on Earth’s surface sweeps across the ground. A given terrestrial user or region remains in coverage for some limited interval. We assume the radius of LEO satellite' beam footprint is $r$. Therefore, a rough coverage time slot $T_{cov}$ can be expressed as $\frac{2r(R+h)}{Rv_s}$. Considering that the computation is mainly done in satellites and their corresponding terrestrial servers, its execution is unaffected by the limitation. Terrestrial users only need to decide whether subscribing the communication service, which can be done in a short time slot. We define the minimum time slot as $T_{min}$. Hence, the maximum velocity $v_{s, max}$ of satellite should satisfy
\begin{equation}
    \begin{split}
        \frac{2r(R+h)}{Rv_{s,max}}\geq T_{min}.
    \end{split}
\end{equation}
Besides, maintaining a minimum SINR threshold is crucial to ensure at least one user can lease the spectrum. If velocity is too high, SINR may drop below the required budget, making spectrum leasing infeasible. Thus, the minimum SINR $SINR_{min}$ should satisfy
\begin{equation}
    \begin{split}
           \zeta SINR_{min} \geq \mathrm{B}_{min}
    \end{split}
\end{equation}
where $\mathrm{B}_{min}$ denotes the minimum budget of the terrestrial user in the beam footprint. Thus, the maximum velocity $v_{s, max}$ of satellite need to satisfy
\begin{equation}
    \begin{split}
           \frac{\zeta\mathcal{P}_{n}g_{n}(\theta_n)G_{s}(\alpha_n^M)\lambda^2}{(16{\pi}^2{d_{n}}^2f_{n}(\theta_n)I+\upsilon^2)(\frac{v_{s,max}}{C}cos(\gamma)\varpi + 1)} \geq \mathrm{B}_{min}.
    \end{split}
\end{equation}

\emph{Definition 1:} The FL algorithm can achieve the global optimal convergence if it satisfies \cite{FLanalysis1, FLanalysis2}
\begin{equation}
    \begin{split}
          | F(\textbf{w}) - F(\textbf{w}^{\ast}) | \leq \varrho,   
    \end{split}
\end{equation}
where $\varrho$ is a small positive constant $\varrho > 0$.

\emph{Theorem 1:} When $F(\textbf{w})$ is a $\eta$\text{-}convex and $\sigma$\text{-}smooth function, the upper bound of $\left[F(\textbf{w}) + F(\textbf{w}^{\ast})\right]$ can be expressed
\begin{equation}
\begin{split}
        F(\textbf{w}^{\ast}) - F(\textbf{w}^{\ast}) \leq \varrho(F(\textbf{w}(0)) - F(\textbf{w}^{\ast})).
\end{split}
\end{equation}

\emph{Proof}: The details of the proof can be seen in \cite{FLanalysis2, Flanalysis3}. For appropriate selections of the iteration numbers, the FL
algorithm will finally converge to the global optimality (24),
the more proof analysis can be found in \cite{FLanalysis2, Flanalysis3}.

\begin{table}
    \caption{Parameter settings}
    \label{tab:Notations}
    \centering
    \begin{tabular}{|c|c|}
    \hline
         Parameter & Value \\
         \hhline{|=|=|}
         Learning rate $l$ & $0.001$ \\
         \hline
         Probability $\vartheta$ & $0.2$ \\
         \hline
         Batch size & $16$ \\
         \hline
         Discount factor $\gamma$ & $0.95$ \\
         \hline
         Antenna gain of the terrestrial user &$[1, 5]$ dBi\\
         \hline
         Transmit power range of terrestrial user & $[100, 1000]$ mW\\
         \hline
         Antenna gain of LEO satellite& $20$ dBi\\
         \hline
         Background noise & $-174$ dBm/MHz\\
         \hline
         Dopper fading coefficient $\varpi$ & $10^5$\\
         \hline
         Revenue coefficient $\zeta$ & $1$\\
         \hline
    \end{tabular}
\end{table}

\subsection{Security Analysis}
The proposed reputation-based consensus mechanism is proved to defend against the following attacks.

\emph{1) Reputation Manipulation Attack:} An attacker tries to maliciously increase its reputation to increase the possibility of being elected as a validator.

Since the underlying blockchain guarantees that all the reputation commitments will achieve consensus, the attacker cannot propose a reputation commitment arbitrarily. Specifically, we assume a satellite that participates in the blockchain network in aggregation round $t$ has reputation $Rep_t$, and in aggregation round $t+1$, the reputation has changed ${\Delta}Rep$. Since the reputation is based on the behaviors of the terrestrial users, the reputation $Rep_{t+1} = Rep_{t} + {\Delta}Rep$ is proved by reaching a consensus with each terrestrial user based on the difference between the original and actual QoS. It is assumed that most of terrestrial users will maintain integrity in order ro maintain expected QoS. Satellites cannot modify the score during the aggregation round transition. The reputation of each satellite is recorded in the block. 

\emph{2) Sybil Attack:} An attacker tries to create multiple validator identities to gain an unfair advantage in block generation.

Satellites who participate in the blockchain are required to consume a significant amount of reputation score to participate in block validation. This requirement makes it costly for an attacker to create multiple identities, as each would necessitate a substantial reputation score. Besides, since our blockchain is permissioned, new satellites must be admitted via a membership service that authenticates them, thus limiting sybil attacks. Additionally, each satellite’s reputation changes only when legitimate spectrum transactions occur, which requires cooperation with terrestrial users. Colluding satellites that falsify transaction records still risk detection if other honest satellites or users provide contradictory evidence in the blockchain.

\emph{3) Collusion Attack:} A group of satellites coordinates to manipulate the network.

The value of the reward tokens for block generation is intrinsically linked to the network's security and reputation. Any successful attack that undermines the network would likely devalue the token, causing financial losses to the colluding satellites. This inherent risk discourages satellites from attempting collusion.

\begin{figure*}[t!]
    \begin{subfigure}[t]{0.5\linewidth}
            \centering
            \includegraphics[width=3.5in]{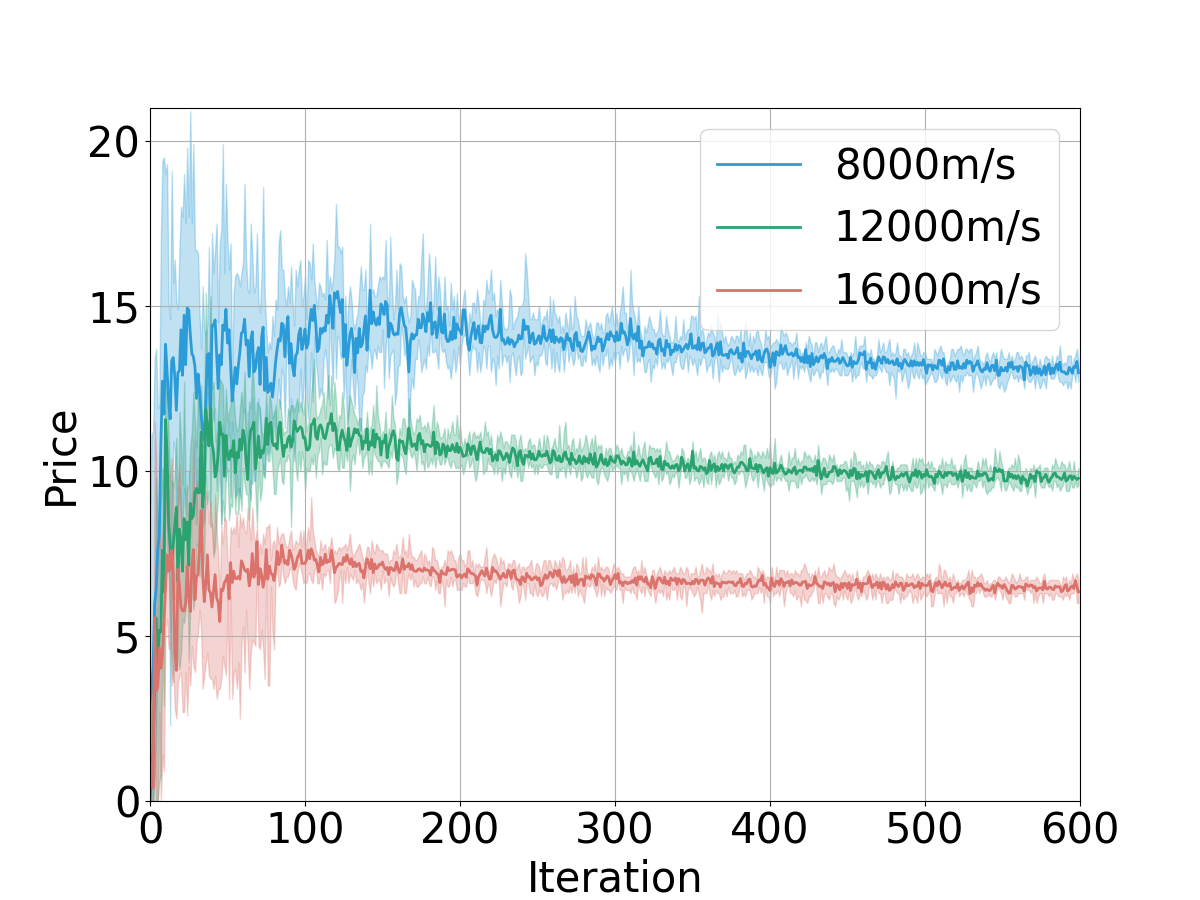}
            \caption{}
            \label{fig:b}
    \end{subfigure}
    \begin{subfigure}[t]{0.5\linewidth}
            \centering
            \includegraphics[width=3.5in]{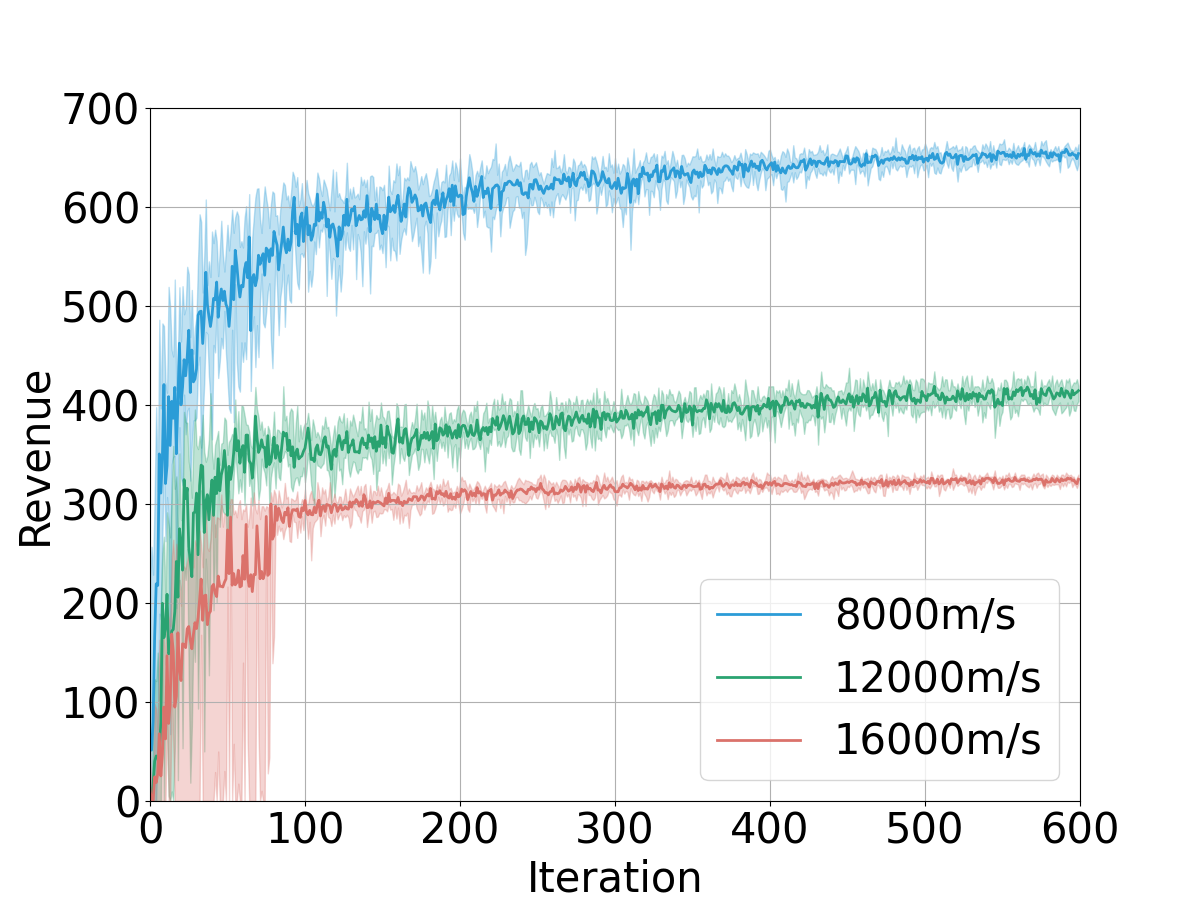}
            \caption{}
            \label{fig:b}
    \end{subfigure}
    \caption{Performance of price and revenue in different velocities of satellites.}
\end{figure*}

\begin{figure*}[t!]
    \begin{subfigure}[t]{0.31\linewidth}
           \centering
           \includegraphics[width=2.35in]{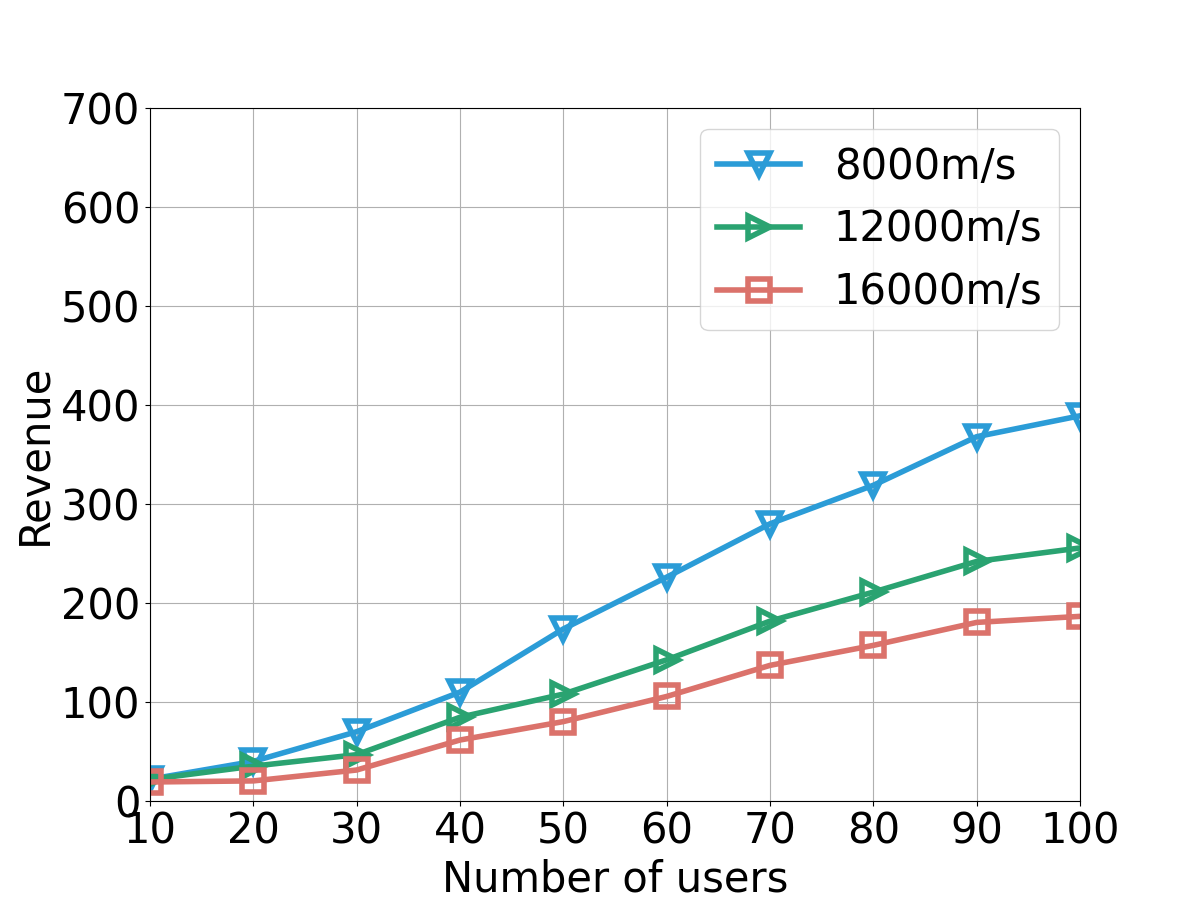}
            \caption{60\% Visibility}
            \label{fig:Case1}
    \end{subfigure}
    \begin{subfigure}[t]{0.31\linewidth}
            \centering
            \includegraphics[width=2.35in]{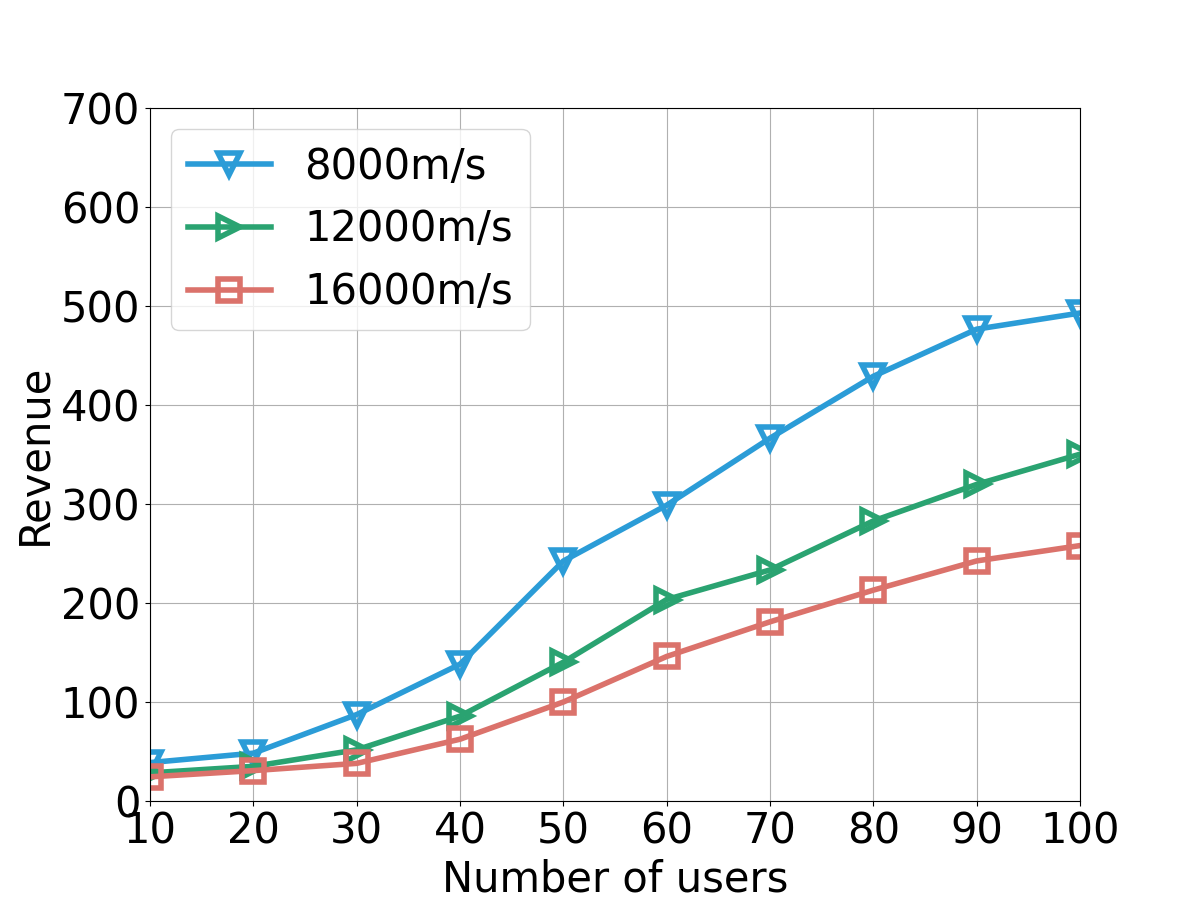}
            \caption{80\% Visibility}
            \label{fig:b}
    \end{subfigure}
    \begin{subfigure}[t]{0.31\linewidth}
            \centering
            \includegraphics[width=2.35in]{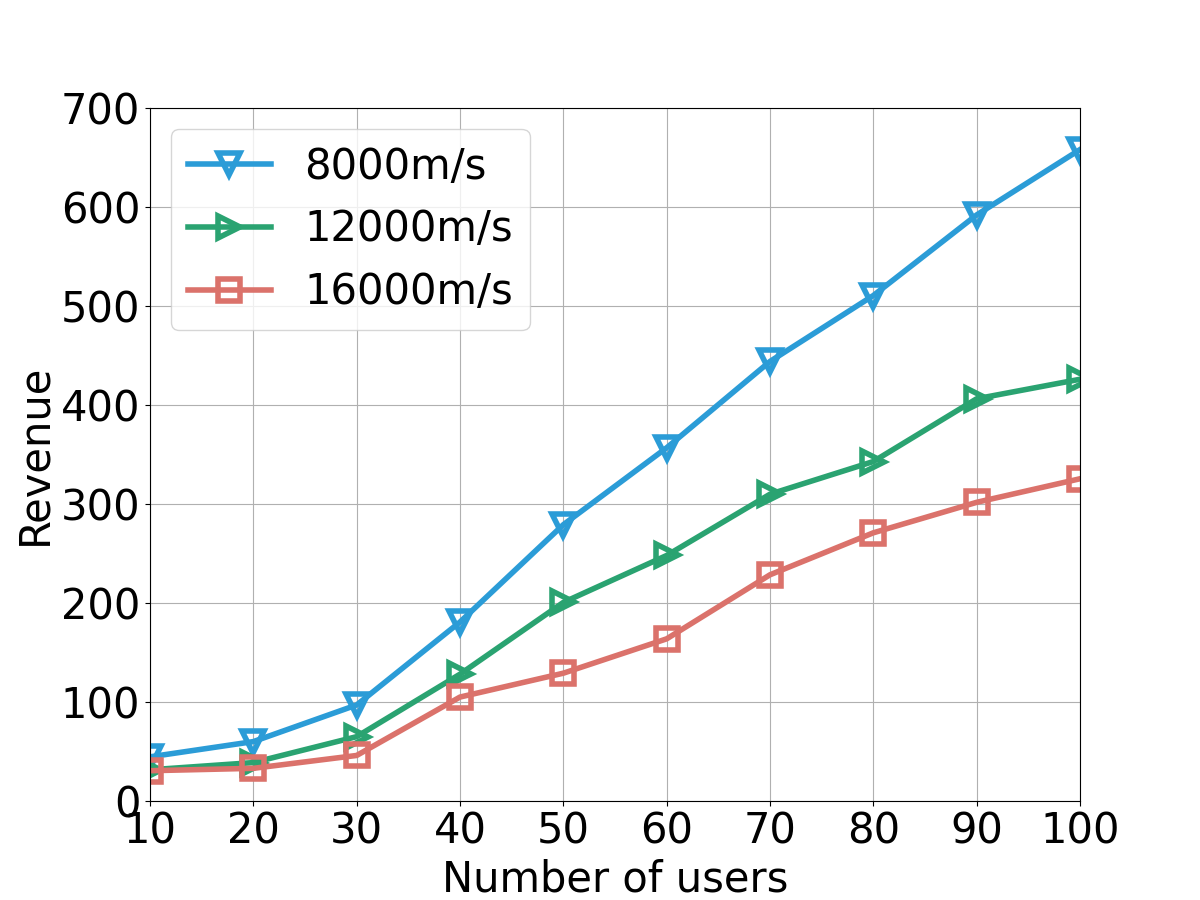}
            \caption{100\% Visibility}
            \label{fig:b}
    \end{subfigure}
    \caption{Average revenue with different velocity and visibility of satellites.}
\end{figure*}

\begin{figure}[!t]
  \begin{center}
  \includegraphics[width=3.5in, height=2.8in]{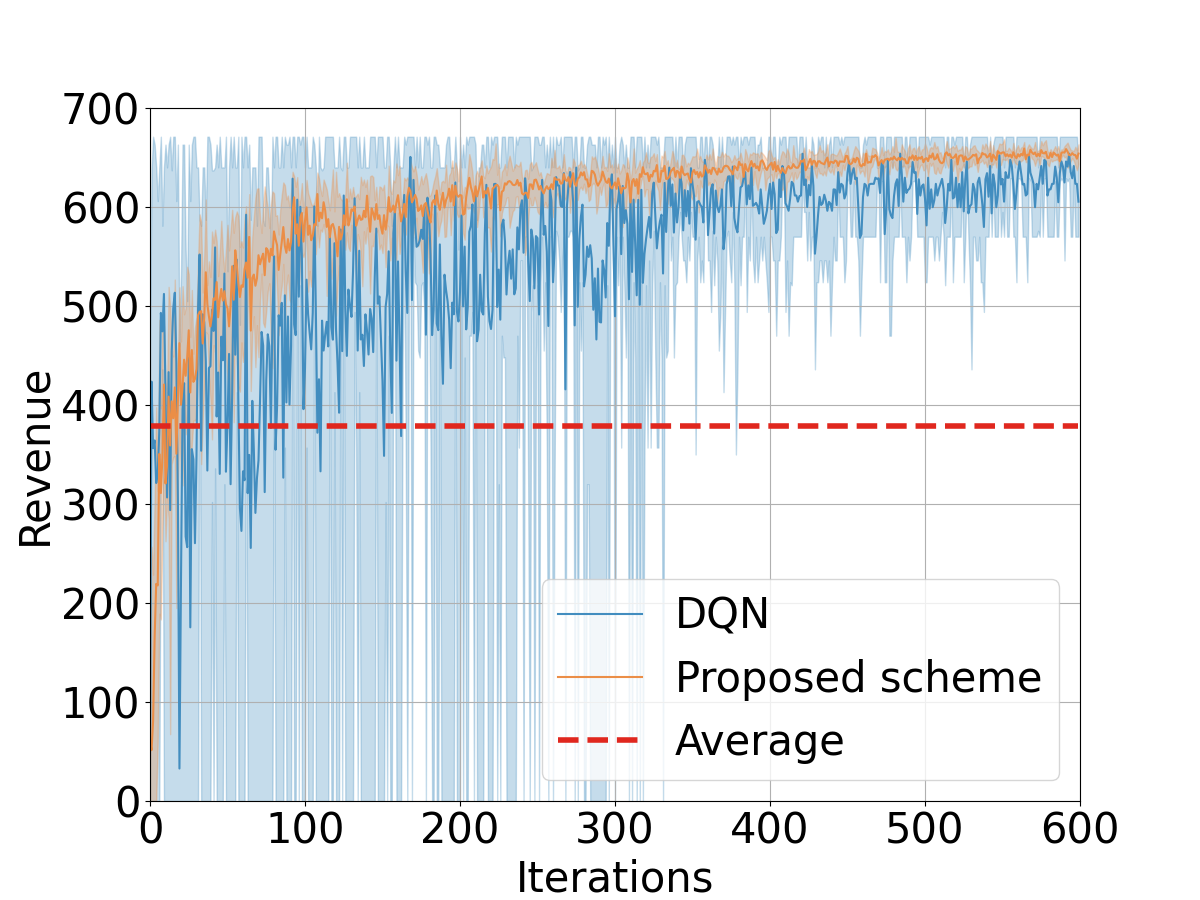}
  \caption{Performance of revenue in different methods }\label{sim_opt_eff_classFinv}
  \end{center}
\end{figure}

\section{Numerical Results}
In this section, simulations are conducted to present the performance of the scheme.

\subsection{Simulation Settings}
We generated multiple terrestrial users who are interested in leasing the LEO satellite spectrum. These users are randomly located at the beam coverage area of the corresponding satellite. The satellite is located at an altitude of $10000$ m above the ground center point. Fig. 6 presents the distribution of terrestrial users and LEO satellites. The blue dots represent terrestrial users and the yellow triangles represent the corresponding LEO satellites in that coverage area. Each terrestrial user generated its potential budget. Besides, we set the learning rate $l$ as 0.001, probability $\vartheta$ as $0.2$, batch size as $16$, antenna gain of the terrestrial user from $1$ dBi to $5$ dBi, transmit power range of terrestrial user from $100$ mW to $1000$ mW, antenna gain of satellite as $20$ dBi, background noise as $-174$ dBm/MHz, Doppler fading coefficient $\varpi$ as $10^5$, revenue coefficient $\zeta$ as $1$. For the model, we employ two linear layers, where the hidden size of each is $16$. The parameter is shown in Table IV.

In the simulations, we used the following metrics to evaluate the algorithm performance:
\begin{itemize}
    \item \textbf{Reward:} The sum of the rewards obtained in each iteration. An increase in the sum of rewards in each iteration indicates that the agent is learning a better policy with the iterations.
    \item \textbf{Price:} The price of idle spectrum of LEO satellites, which is needed to be adjusted by satellite operators to maximize the revenue.
    \item \textbf{Revenue:} The revenue obtained by leasing idle spectrum of satellite to terrestrial users.
\end{itemize}


\subsection{Case Study}

In Fig. 7, we conducted the simulation to show the performance of average reward, price, and revenue with different numbers of FL satellite agents. Specifically, Fig. 7(a), Fig. 7(b) and Fig. 7(c) are in the case where the number of participating agents is 10, Fig. 7(d), Fig. 7(e) and Fig. 7(f) are in the case where the number of participating agents is 20, and Fig. 7(g), Fig. 7(h) and Fig. 7(f) are in the case where the number of participating agents is 30. We set the number of terrestrial users as $100$ and the velocity of the LEO satellite as $8000$ m/s. First, it can be observed that the variance of the maximum value and the minimum value in the figures decreases with the increase in the number of participating agents. This is because the global model can be influenced by those local model parameters uploaded by agents who do not learn a good policy, more participating agents can mitigate the negative influence. Besides, it can be observed that large variances of performance occur after about 50 iterations in the case where the participating agents are 20 and 30, and the variances decrease after about 150 iterations. This is because some agents may not learn a better policy in the process of model training, thus reducing the average performance of the network. But the performance of those agents who did no learn optimal policy increases after more rounds of iterations due to the effect of FL.

In Fig. 8, we mainly study the influence of LEO satellite's movements. 100 terrestrial users are generated for simulation. We compared the performance in the scenarios where the relative velocity of LEO satellites are $8000$ m/s, $12000$ m/s and $16000$ m/s. We can observe that the price reaches about $14$, $10$ and $6$ respectively, and the revenue reaches about $650$, $410$ and $310$ respectively after 600 iterations when the relative velocity of LEO satellite is $8000$ m/s, $12000$ m/s and $16000$ m/s.  Generally, it can be observed that the price of the spectrum and the revenue decrease with the increase in the relative velocity of LEO satellite. This is because the spectrum is influenced by the Doppler shift, the QoS of terrestrial users who use the channel decreases, which leads to the decrease in the price and revenue.

In Fig. 9, we compare the revenue in different numbers of users with different visibility and different satellite velocities. We take the weather condition into consideration in cases. Thus, We define the visibility as the percentage of time that terrestrial user can use the spectrum without losing connection by the influence of bad weather conditions. In the case where visibility is $60\%$, the revenue increases from about $20$ to about $400$, $250$ and $200$ with the increase of the number of users when the relative velocity is $8000$ m/s, $12000$ m/s and $16000$ m/s. When the visibility is $80\%$, the revenue increases from about $20$ to about $500$, $350$ and $260$, and when the relative is $100\%$, the revenue increases from about $20$ to about $660$, $410$ and $320$. It can be observed that a lower visibility reduces the revenue. This is because low visibility leads to low quality of spectrum. This impact makes the price decrease, thus decreasing the revenue.

In Fig. 10, we compare the performance of three different methods. The yellow line is the performance of the proposed scheme. It can be observed that the revenue has achieved a significant level after 150 iterations. The red line is the method by which the operators set the average budgets of terrestrial users as the price of spectrum. We can observe that the revenue is about $390$, which is quite lower than the proposed scheme. The blue line is the method which uses DDQN to find optimal policy. It can be observed that although the agent may get the optimal policy, the variance of the performance is too large compared to the proposed scheme. Besides, the convergence speed of the proposed scheme is faster than the DDQN algorithm.

\section{Conclusion}
In this paper, we consider the effective spectrum pricing and uplink transmit power control scheme for LEO satellite IoT. We first formulate a reinforcement problem based on the satellite communications features to maximize the benefits of leasing spectrum. Next, a locally trained DRL-based scheme is proposed for satellites to find the optimal policy. Then, we further introduce a blockchain-driven FL framework to enhance the training collaboration while keeping the system distributed throughout the whole process to guarantee the security of local private information. We also conduct simulations to present the pricing performances of agents that participate in the FL and compare the performance of the learning-based scheme and the non-learning-based scheme. Numerical results show the efficiency of the spectrum pricing and power control strategy proposed in this paper. In the process of LEO satellite idle spectrum leasing, terrestrial users may move to another area while still leasing the previous spectrum or there may be a sudden surge or decrease of users in that area after leasing a certain spectrum. In this case, such problems may arise: 1) The QoS obtained by the LEO satellite at this time may vary, such as changes in interference and power attenuation due to different numbers of accesses and distance values. 2) The latest price of that spectrum may fluctuate. Therefore, in our future work, we will focus on the spectrum allocation in a receptive and timely manner while maintaining the pricing in an acceptable range.

\bibliographystyle{ieeetr}
\bibliography{reference}

\ifCLASSOPTIONcaptionsoff
  \newpage
\fi

\vfill

\end{document}